\documentclass[twocolumn,10pt,longbibliography,superscriptaddress,reprint,amsmath,amssymb,aps,nofootinbib]{revtex4-2}

\usepackage{textcomp}
\usepackage{amsfonts}
\usepackage{calc}
\usepackage{mathrsfs}
\usepackage{amssymb}
\usepackage{amsmath}
\usepackage{array}
\usepackage{tensor}
\usepackage{color}
\usepackage{bm}
\usepackage{graphicx}
\usepackage{hyperref}
\usepackage{cleveref}
\usepackage{booktabs}
\usepackage{float}
\usepackage[scr=boondoxo,scrscaled=1.05]{mathalfa}
\usepackage[dvipsnames]{xcolor}
\usepackage[normalem]{ulem}
\usepackage{orcidlink}
\usepackage{yfonts}

\newcommand{\e}{\varepsilon}

\newcommand{\iom}{i \omega}
\newcommand{\ptw}{\psi_2}
\newcommand{\dptw}{\psi_2'}
\newcommand{\po}{\psi_1}
\newcommand{\dpo}{\psi_1'}
\newcommand{\pz}{\psi_0}
\newcommand{\dpz}{\psi_0'}
\newcommand{\pzbf}{\psi_{0b}+M_{2af}}
\newcommand{\dpzbf}{\psi_{0b}'+M_{2af}'}
\newcommand{\nn}{\nonumber}

\newcommand{\h}{h_{\alpha \beta}^{1}}
\newcommand{\hL}{h_{\alpha \beta}^{1\text{L}}}
\newcommand{\hLr}{h_{\alpha \beta, r_0}^{1\text{L}}}

\newcommand{\hRW}{h_{\alpha \beta}^{1\text{RW}}}
\newcommand{\mtaf}{M_{2af}}%{\varphi_0}
\newcommand{\mtafr}{M_{2af,r_0}}

\newcommand{\rp}{r_{\mathrm{p}}}
\newcommand{\phip}{\phi_{\mathrm{p}}}
\newcommand{\s}{{\tilde{t}}}

\newcommand{\dpsi}{\phi}

\newcommand{\eqn}[1]{Eq.~(\ref{#1})}

\emergencystretch=\maxdimen
\begin{document}
	
	\preprint{APS/123-QED}
	
	\title{Slow evolution of the metric perturbation due to a quasicircular inspiral into a Schwarzschild black hole}
	
	\author{Leanne Durkan\,\orcidlink{0000-0001-8593-5793},}
	\author{Niels Warburton\,\orcidlink{0000-0003-0914-8645},}
	\affiliation{School of Mathematics and Statistics, University College Dublin, Belfield, Dublin 4, Ireland}
	
	\date{\today}
	
	\begin{abstract}		
	
	Extreme mass-ratio inspirals (EMRIs) are one of the most highly anticipated sources of gravitational radiation novel to detection by millihertz space-based detectors. 
	To accurately estimate the parameters of EMRIs and perform precision tests of general relativity, their models should incorporate self-force theory through second-order in the small mass ratio. 
	Due to their extreme mass ratio, EMRIs inspiral slowly when sufficiently far from merger, and a two-timescale approximation can be applied.
	Within the two-timescale approach, the slow evolution of the first-order metric perturbation contributes to the source for the second-order metric perturbation, and must be accounted for in EMRI waveform models. 
	In this paper we calculate the slow evolution of the first-order metric perturbation in the Lorenz gauge for quasicircular orbits on a Schwarzschild background in the frequency domain. 
	Lorenz gauge solutions to the first-order metric perturbation and its slow evolution are obtained via a gauge transformation from Regge-Wheeler gauge solutions. The slow evolution of Regge-Wheeler and Zerilli master functions, in addition to a gauge field are determined using the method of partial annihilators. 		
	\end{abstract}
	
	\maketitle

	\section{Introduction}\label{sec:intro}
	
	Gravitational wave (GW) astronomy has seen huge progress since the first discovery by the LIGO/Virgo Collaboration in 2015 \cite{PhysRevLett.116.061102}. 
	To date, GWs have been detected from numerous sources including compact binaries with mass ratios from 1:1 to $\sim$26:1  \cite{LIGOScientific:2021djp}, binary neutron stars \cite{PhysRevLett.119.161101} and black hole-neutron star mergers \cite{LIGOScientific:2021qlt}. 
	The next generation of space-based GW detectors such as the Laser Interferometer Space Antenna (LISA) \cite{LISA:2017pwj}, with access to millihertz frequencies, will expand the current parameter space of compact binaries available to detection. 
	This drives the need to develop GW models for millihertz sources: extreme-mass-ratio inspirals (EMRIs), one of the key anticipated sources of GWs detectable by LISA. 
	EMRIs are binary systems of compact objects in which the larger body (the primary) has a mass $M$ that is at least $10^4$ times that of the smaller body (the secondary) with mass $\mu$. 
	Astrophysical observations establish the primary as a supermassive black hole with a mass of $\sim 10^4 - 10^7 M_\odot$ residing in galactic centres, with the secondary a stellar-mass compact object, either a black hole (BH), neutron star, or some exotic compact object \cite{Berry:2019wgg}. 
	EMRI GW frequencies range between $\sim 10^{-3} - 10^{-2}$ Hz \cite{LISA:2017pwj}, placing them comfortably within the LISA band, detectable so long as their signal is sufficiently loud \cite{Babak:2017tow}. 
	It is estimated that LISA will observe between a few and a few thousand EMRIs over its lifetime \cite{Babak:2017tow}. 
	This estimate provides substantial motivation to develop models of EMRIs with which to perform matched filtering of LISA's future data stream. 
	
	Black hole perturbation theory (BHPT) is a natural choice for modelling EMRIs. 
	The metric of the binary, $g_{\alpha \beta}$, is expanded around the metric of the primary,  $g^{0}_{\alpha \beta}$, in powers of the mass ratio $\e = \mu/M \ll 1$ \cite{Miller:2020bft}:
	\begin{equation}\label{eq:metric_expansion}
		g_{\alpha \beta} = g^0_{\alpha \beta}\left(x^{\mu}\right)+\sum_{n=0}^\infty \e^{n} h_{\alpha \beta}^{n}\left(x^{\mu} ; z^{\mu}\right),
	\end{equation}
	where $g^{0}_{\alpha \beta}$ depends on the background coordinates $x^\mu$, which describe the primary, and $h^{n}_{\alpha \beta}$ is the $n^{th}$-order metric perturbation, depending on both the background coordinates and the position of the secondary's world-line, $z^\mu(\tau)$. 
	%We will drop the explicit dependence of $g^{0}_{\alpha \beta}$ and $h_{\alpha \beta}^{n}$ on $x^\mu$ and $z^{\mu}$ hereafter. 
	In this paper we take $g^0_{\alpha\beta}$ to be the Schwarzschild metric, with coordinates $x^\mu = \{t,r,\theta, \phi\}$ and without loss of generality we constrain the secondary to the equatorial plane such that $z^\mu(\tau) = \{t_{\mathrm{p}},\rp,\pi/2, \phip\}$.
	
	% \NW{\begin{itemize}
	% 	\item Say phase accuracy is important
	% 	\item Say two-timescale shows we need to go to second-order in the mass ratio
	% 	\item Specialize to circular orbits and give the two-timescale expansion
	% 	\item Say how t derivatives are handled
	% 	\item Say how the $\tilde{t}$ derivatives are handled
	% 	\item Introduce operators and show the usual method gives a non-compact source
	% 	\item say even worse in the Lorenz gauge as we need to solve for all the coupled modes
	% 	\item say our method efficiently deals with both problems.
	% \end{itemize}
	% }
	
	Searching for and parameterizing EMRI waveforms in the LISA data stream relies crucially on theoretical waveform templates, where the phase error of the template with respect to the true signal must be $\ll 1$ radian \cite{LISA:2017pwj,Hinderer:2008dm, Miller:2020bft}.
	In order to reach this criteria the expansion in Eq.~\eqref{eq:metric_expansion} must be carried through second-order in the mass ratio \cite{Hinderer:2008dm}.
	
	Due to their small mass ratio, orbital parameters such as the radius, orbital energy, angular momentum, frequency and amplitude of EMRIs evolve slowly during the inspiral \cite{Miller:2020bft}. That is, they evolve on the radiation-reaction timescale.
	On the other hand, the orbital phase evolves on the orbital timescale, and accumulates rapidly over many tens or hundreds of thousands of orbits \cite{Hinderer:2008dm}.
	The presence of these disparate timescales allows for a two-timescale expansion where at each order in $\e$ the metric perturbation is written as a product of slowly evolving amplitudes and a rapidly evolving phase.
	Hereafter we specialize to quasicircular inspirals and define the ``fast time'' $t$ such that $\phip(t) = \int^t \Omega(\varepsilon t^\prime)d t^\prime$, where $\Omega$ is the orbital frequency. We then define the ``slow time'' to be $\tilde{t} = \e t$ \cite{Miller:2020bft}.
	At each order we write the metric perturbation as:
	\begin{align}\label{eq:twotimescale_expansion}
		h^n_{\alpha\beta}(\tilde{t},\rp,\phi_p,x_i) = \sum_{m\in \mathbb{Z}} h^{n,m}_{\alpha\beta}(\tilde{t},\rp,x^i)e^{-im\phi_p},
	\end{align}
	where $x_i = (r,\theta,\phi)$.
	The field equations for the metric amplitudes $h^{n,m}_{\alpha\beta}$ can be derived by substituting Eqs.~\eqref{eq:metric_expansion} and \eqref{eq:twotimescale_expansion} into the Einstein Field Equations and expanding order-by-order in $\e$.
	The field equations must also be regularized in order to compute the regular contribution to the metric perturbation that gives rise to the self-force (SF) that drives the inspiral \cite{Poisson:2011nh,Barack:2018yvs}.
	Through second-order in the mass ratio, this regularization procedure is presently best understood in the Lorenz gauge \cite{Pound:2012nt,Pound:2014xva}.
	
	Using the two-timescale expansion, time derivatives that appear in the Einstein tensor can be handled using $d\phip/dt = \Omega$ and $d\tilde{t}/dt = \e$, such that:
	\begin{align}
		\partial_t = \Omega \partial_{\phip} + \e \partial_{\tilde{t}}.
	\end{align}
	The presence of the second term means there is a contribution from the ``slow-time derivative'' of the first-order metric perturbation to the source of the second-order metric perturbation \cite{Miller:2020bft}.
	The explicit form of the radial field equations after further decomposing the metric perturbations onto a basis of tensor spherical harmonics and applying the Lorenz gauge condition can be found in Eqs.~(16) and (17) of Ref.~\cite{Miller:2020bft}.
		
	The goal of this paper is to compute $\partial_{\s} h^1_{\alpha\beta}$, as it contributes to the source for the second-order metric perturbation and hence is a necessary ingredient for accurate EMRI waveform templates. 
	We begin by expanding the orbital radius and frequency as $\rp(\s) = r_0(\s) + \mathcal{O}(\e)$ and $\Omega(\s) = \Omega_0(\s) + \mathcal{O}(\e)$ where $r_0$ is the radius of a circular geodesic with orbital frequency $\Omega_0 = \sqrt{M/r_0^3}$.
	We then write the slow-time derivative of the first-order metric perturbation as\footnote{There are also contributions from the slow evolution of the mass and spin of the primary which we ignore here as they can be handled separately in a straightforward manner \cite{Miller:2020bft}.}:
	\begin{equation}
		\partial_{\tilde{t}}h_{\alpha \beta}^1 = \dfrac{d r_0}{d\tilde{t}}~\partial_{r_0} h_{\alpha \beta}^1 + \mathcal{O}(\e). \label{eq:slowtimehrad}
	\end{equation}
	The first term on the right hand side can be computed using $dr_0/d\s = (dE/dr_0)^{-1} dE/d\s$, with the balance law $dE/d\s = -\mathcal{F}$, where $E$ is the (specific) orbital energy and $\mathcal{F}$ is the sum of the radiated energy flux of gravitational waves through the event horizon and null infinity.
	The computation of the slow-time derivative of $h^1_{\alpha\beta}$ is thus converted to computing the  $r_0$-derivative of the first-order metric perturbation amplitude.
	
	\subsection{Lorenz gauge metric perturbations}
	
	We can decompose the trace-reversed metric perturbation $\bar{h}_{\alpha\beta}^1 = h_{\alpha\beta}^1 - \tfrac{1}{2}g_{\alpha\beta}h^1$ into spherical harmonic and Fourier modes (dropping the dependence on $\rp$ for convenience): 
	\begin{align}\label{eq:LorenzFourierDecomp}
		\bar{h}_{\alpha\beta}^1 = \sum_{l=0}^{\infty} \sum_{m=-l}^{l}\sum_{i=1}^{10} \frac{a^{(i)}_l}{r}\bar{h}^{(i)}_{l m}(r) Y_{\alpha\beta}^{(i)l m}(r,\theta,\phi)e^{-i m \Omega_0 t},
	\end{align}
	where the $Y_{\alpha\beta}^{(i)l m}$ are the Barack-Sago-Lousto (BSL) tensor spherical harmonic basis \cite{Barack:2005nr,Barack:2007tm} and:
	\begin{equation}
	a^{(i)}_{l}=\frac{1}{\sqrt{2}} \times \begin{cases}1 & \text { for } i=1,2,3,6, \\ 1 / \sqrt{l(l+1)} & \text { for } i=4,5,8,9, \\ 1 / \sqrt{(l+2)(l+1)l(l-1)} & \text { for } i=7,10.\end{cases}
	\end{equation}
	Substituting this decomposition into the linearized Einstein field equations and applying the Lorenz gauge condition $\nabla_\alpha \bar{h}^{\alpha\beta} = 0 $ we arrive at the radial field equations \cite{Barack:2005nr,Barack:2007tm,Akcay:2010dx,Akcay:2013wfa,Wardell:2015ada}:
	\begin{equation}\label{eq:LorenzGaugeRadial}
		\square_{lm} \bar{h}^{(i)}_{l m} - 4 f^{-2}\mathcal{M}^{(i)}_{\;\;(j)} \bar{h}^{(j)}_{l m} = \mathcal{J}^{(i)}_{l m},
	\end{equation}
	where $\square_{lm}=\partial_r^2 + f'/f \partial_r - f^{-2}[V_l(r) - \omega_m^2]$ with $f(r) = 1- 2M/r$, $V_l(r) = f[2M/r^3 + l(l+1)/r^2]$, $\omega_m = m \Omega_0$, and $\mathcal{M}^{(i)}_{\;\;(j)}$ is a matrix that encodes the coupling between the various components of the metric perturbation, whose components can be found explicitly in \cite{Barack:2005nr,Barack:2007tm}, and $\mathcal{J}^{(i)}_{l m} \propto \delta(r - r_0)$, is the stress-energy tensor describing the source of the perturbation, as decomposed onto this basis. Taking an $r_0$-derivative of the radial field equations we get:
	\begin{equation}\label{eq:r0-deriv_Lorenz}
		\square_{lm} \bar{h}^{(i)}_{l m,r_0} - 4 f^{-2}\mathcal{M}^{(i)}_{\;\;(j)} \bar{h}^{(j)}_{l m,r_0} = \mathcal{J}^{(i)}_{l m,r_0} - \square_{lm,r_0} \bar{h}^{(i)}_{l m} .
	\end{equation}
	This equation is challenging to solve for two main reasons: (i) the source on the right-hand side is unbounded, and (ii) the coupling between the tensor harmonic $i$-modes via the $\mathcal{M}^{(i)}_{(j)}$ matrix increases the complexity and computational burden of the calculation.
	The non-compact source is particularly challenging for the standard variation of parameters approach for constructing inhomogeneous solutions to ordinary differential equations (ODEs), as this approach relies on having the homogeneous solutions computed for all radii inside the source.
Nonetheless a numerical code using the variations of parameters method to find solutions to Eq.~\eqref{eq:r0-deriv_Lorenz} has been implemented in Ref.~\cite{Miller_etal}.
	
	We present in this paper a novel approach to calculating $\partial_{\tilde{t}} h^{1}_{\alpha\beta}$ by (i) making use of the a gauge transformation from Regge-Wheeler (RW) to Lorenz gauge solutions \cite{Berndtson:2007gsc, Hopper:2012ty} and (ii) using the method of partial annihilators \cite{Hopper:2012ty} to replace the unbounded source with a compact one, at the expense of introducing a higher-order operator on the left hand side of the field equations.
	The combination of these two techniques means we only have to numerically solve homogeneous equations for a handful of uncoupled scalar fields that describe the RW master variables and their $r_0$ derivatives. We can then construct the inhomogeneous solutions entirely from data obtained on the world-line.
	We find this approach more efficient and easier to implement than the variation of parameters method with an unbounded source, such as that of Ref.~\cite{Miller_etal}.
	
	\subsection{Outline of our approach}
	
Our method is outlined as follows. 
For a particular choice of gauge vector $\xi^{\alpha}\sim \mathcal{O}(\varepsilon)$, the infinitesimal coordinate transformation: $x^{\alpha} \rightarrow x'^{\alpha} = x^{\alpha} + \xi^{\alpha} $ allows us to write the gauge transformation of the metric perturbation to leading order as:
\begin{equation}
	h^{1 \text{L}}_{\alpha \beta}=h^{1 \text{RW}}_{\alpha \beta} + \pounds_{\xi} g^{0}_{\alpha \beta}, \label{gaugetransformeqn}
\end{equation} 
where $\hL$ and $\hRW$ are the metric perturbation in the Lorenz gauge and RW gauge respectively, and $\pounds_{\xi}$ is the Lie derivative with respect to the gauge vector $\xi^{\alpha}$.
The right-hand side of \eqn{gaugetransformeqn} can then be written in terms of solutions to the spin-weighted RW and Zerilli (RWZ) master equations.
Taking the derivative of \eqn{gaugetransformeqn} with respect to $r_0$, we obtain $\partial_{r_0} \hL$ in terms of $\partial_{r_0} \hRW$ and the gauge vector contribution: $-2\partial_{r_0}\xi_{(\alpha;\beta)}$.
Numerically, it is significantly easier to solve for the RWZ master functions and their $r_0$-derivatives then apply the gauge transformation from \eqn{gaugetransformeqn} than to solve Eq.~\eqref{eq:r0-deriv_Lorenz} directly to obtain $\partial_{r_0} \hL$. 
Expressions for the 10 independent components in \eqn{gaugetransformeqn} were derived by Berndtson \cite{Berndtson:2007gsc}. The homogeneous expressions are provided explicitly in this paper in \Cref{parital annihilators} and \Cref{BEven} for the case where $l \geq 2$ and $\omega_m \neq 0$. We shall drop the subscript $m$ on $\omega$ henceforth.

The use of the gauge transformation in \eqn{gaugetransformeqn} and its $r_0$-derivative simplifies the calculation of $\partial_{r_0}h^{1\text{L}}_{\alpha\beta}$, but we are still left with the task of computing the $r_0$-derivative of the RWZ master functions.
As with the Lorenz gauge equations, the radial equation for the $r_0$-derivative of the RWZ master functions will have a source with unbounded support.
To simplify the calculation we employ the method of partial annihilators which we illustrate now with a toy example.
Consider an inhomogeneous second-order ODE of the form:
\begin{align}\label{eq:example_ODE}
	\mathcal{L} \varphi = S,
\end{align}
where $\mathcal{L} \equiv \partial^2_r + \omega(r_0)^2$, $\varphi = \varphi(r;r_0)$ and $S = \mu \delta(r-r_0)$.
Differentiating \eqn{eq:example_ODE} with respect to $r_0$ we arrive at:
\begin{align}\label{eq:example_ODE_r0-deriv}
	\mathcal{L} \varphi_{,r_0} = S_{,r_0} - 2\omega\omega_{,r_0} \varphi,
\end{align}
which, just like Eq.~\eqref{eq:r0-deriv_Lorenz}, has a source which is unbounded due to the presence of $\varphi$ in the source, which is defined $ \forall ~r$.
If we now act on \eqn{eq:example_ODE_r0-deriv} with $\mathcal{L}$ and make use of Eq.~\eqref{eq:example_ODE} we get:
\begin{align}
	\mathcal{L}^2 \varphi_{,r_0} = \mathcal{L} S_{,r_0} - 2\omega\omega_{,r_0} S.
\end{align}
We now have a fourth-order operator, $\mathcal{L}^2$, on the left hand side and a compact source on the right, containing only Dirac delta functions and their derivatives.
We find equations of this form much easier to solve numerically than equations of the form of Eq.~\eqref{eq:example_ODE_r0-deriv}.
As such we will use this technique multiple times throughout this work, including cases where we have to introduce a sixth-order operator on the left-hand side in order to arrive at a compact source.

	In this paper we will focus on radiative modes as the static modes are known analytically \cite{Osburn:2014hoa} and thus it is straightforward to take the $r_0$ derivative of them.
	The rest of this paper proceeds as follows. 
	In \Cref{RW} we introduce the RW gauge by writing the metric perturbation in terms of a tensor spherical harmonic basis and Fourier decomposition. 
	We introduce the spin-weighted RWZ master equations and describe how to obtain their inhomogeneous solutions using the method of variation of parameters in \Cref{psiretsec}. 
	We find their homogeneous solutions using the \texttt{ReggeWheeler} \textit{Mathematica} package from the Black Hole Perturbation Toolkit (BHPToolkit) \cite{BHPToolkit}.
	In \Cref{psirretsec} we then use the method of partial annihilators to solve for the $r_0$ derivatives of the RWZ master functions. 
	To compute the homogeneous solutions to the fourth-order equations, we construct appropriate boundary conditions at large radius and near the horizon, and numerically integrate to $r_0$ in \Cref{sec:RWZ_BCs}. In Sec.~\ref{sec:RW_r0deriv_Flux} we show how to calculate the $r_0$ derivative of the GW energy flux at infinity as a demonstration of our method. 
	In Sec.~\ref{sec:RW_results} we provide our numerical results, consistency checks and directly compute the $r_0$ derivative of the GW flux to infinity.

 	In \Cref{parital annihilators} we discuss the Lorenz gauge, providing explicit expressions for $\hL$ in the odd-sector using the gauge transformation derived by Berndtson \cite{Berndtson:2007gsc} for the case where $l \geq 2$ and $\omega \neq 0 $. Berndtson's transformation is equivalent to the transformation in \cite{Hopper:2012ty}.
	The majority of the even-sector expressions are lengthy and are therefore presented in \Cref{BEven}. 
	We note here that the individual components of \eqn{gaugetransformeqn} can also be derived using the recent work of Dolan, Wardell and Kavanagh \cite{Dolan:2021ijg,me}, at least in the homogeneous case.  In the odd-sector, $\hL$ is constructed from two RWZ master functions with compact sources.
	The $r_0$ derivative of each of these obeys an equation of the form of Eq.~\eqref{eq:example_ODE} and as such the method of partial annihilators proceeds similarly to Eq.~\eqref{eq:example_ODE_r0-deriv}. In the even-sector, $\hL$ is constructed from four RWZ master functions, three of which have compact sources and an additional gauge field whose source is unbounded.
	It turns out this additional gauge field, known as $\mtaf$, can be again be tackled using the method of partial annihilators. Solving for $\partial_{r_0}\mtaf$ is a little more involved than our earlier toy example however, and we must solve a pair of sixth-order ODEs with compact sources, discussed in \Cref{mtafrsec}. 
	Solutions to $\partial_{r_0}\hL$ can then be determined by taking the $r_0$ derivative of the gauge transformation in \eqn{gaugetransformeqn} and substituting RWZ master functions and their $r_0$ derivatives, calculated in \Cref{RW}. We also provide the transformation from Berndtson's expressions to the BSL spherical harmonic basis \cite{Barack:2005nr, Barack:2007tm} in \Cref{sec:BSL}, which is commonly used in SF calculations \cite{Barack:2007tm,Barack:2010tm,Akcay:2010dx,Akcay:2013wfa,Pound:2019lzj}.
 	We present our final numerical results and consistency checks for $\hL$ in Sec.~\ref{sec:Lorenz_results}.
	Future prospects are discussed in \Cref{conclusion}. 
	Our calculation of $\partial_{r_0}\hL$ provides an important ingredient for second-order SF calculations and our results have already been used in Refs.~\cite{Warburton:2021kwk} and \cite{Wardell:2021fyy} to calculate the second-order flux and post-adiabatic waveforms, respectively.

	 Raising and lowering indices and differential operations are performed with respect to $g_{\alpha \beta}^0$. 
	 Geometric units will be used throughout this paper such that $G=c=1$, with metric signature $(-+++)$.
		
	\section{Regge-Wheeler gauge}\label{RW}

	In this section we will review the RW gauge and obtain solutions to the RWZ master equations via the method of variation of parameters.
	We then discuss in detail our approach to solve for the $r_0$ derivatives of the RWZ master functions. 
	To define the RW gauge, we must first introduce the spherical harmonic mode and Fourier decomposition of the first-order metric perturbation. 
	The spacetime manifold is chosen such that it is foliated by hypersurfaces defined by constant $t$. 
	In the frequency domain, using a tensor spherical harmonic basis, $h^1_{\alpha \beta}$ can then be written as:
	\begin{equation}
		h^1_{\alpha \beta}= \sum_{l=0}^{\infty} \sum_{m=-l}^{l}   h^{l m}_{\alpha \beta} (r, \theta,\phi) e^{-i \omega t}. \label{metricmodaldecomp}
	\end{equation}
	%Having specialised to quasicircular orbits, we can drop the explicit dependence on $\omega$ henceforth. 
	We can split $h^{l m}_{\alpha \beta}$ into two sectors, with either odd or even parity. 
	The odd-(even-) sector is defined for $l+m$ odd (even) such that:
	\begin{equation}
		h^{l m}_{\alpha \beta}(r,\theta, \phi)=h_{\alpha \beta}^{o, l m}(r,\theta, \phi)+h_{\alpha \beta}^{e, l m}(r,\theta, \phi), \label{hoddeven}
	\end{equation}
	where $h_{\alpha \beta}^{o, l m}$ and $h_{\alpha \beta}^{e, l m}$ are the odd- and even-sector perturbations respectively. Following Berndtson's notation \cite{Berndtson:2007gsc}, the odd-sector metric perturbation contains 3 degrees of freedom given by $h_0^{l m}(r)$, $h_1^{l m}(r)$ and $h_2^{l m}(r)$. For convenience we suppress the explicit dependence on $r$ of the odd-sector fields henceforth. The full odd-sector metric perturbation is then written as \cite{Berndtson:2007gsc}:
	\begin{align}
	&h_{\mu \nu}^{o, l m}=\left(\begin{array}{cccc}
	{0} & {0} & {h_{0}^{l m} \csc \theta \frac{\partial Y_{l m}}{\partial \phi}} & {-h_{0}^{l m} \sin \theta \frac{\partial Y_{l m}}{\partial \theta}} \\
	{*} & {0} & {h_{1}^{l m}\csc \theta \frac{\partial Y_{l m}}{\partial \phi}} & {-h_{1}^{l m} \sin \theta \frac{\partial Y_{l m}}{\partial \theta}} \\
	{*} & {*}& {-h_{2}^{l m}X_{l m}} & {h_{2}^{l m}\sin \theta W_{l m}} \\
	{*} & {*} &{*} & {h_{2}^{l m} \sin ^{2} \theta X_{l m}}\label{oddperturbation}
	\end{array}\right),
	\end{align}
	where $X_{lm}$ and $W_{lm}$ are defined by \cite{Berndtson:2007gsc}: 
	\begin{align}
		W_{l m}(\theta, \phi) = &\frac{\partial^{2} Y_{l m}}{\partial \theta^{2}}-\cot \theta \frac{\partial Y_{l m}}{\partial \theta} -\frac{1}{\sin ^{2} \theta} \frac{\partial^{2} Y_{l m}}{\partial \phi^{2}}, \label{eq:Wlm} \\
		X_{l m}(\theta, \phi) = &\frac{2}{\sin \theta} \frac{\partial}{\partial \phi}\left(\frac{\partial Y_{l m}}{\partial \theta}-\cot \theta Y_{l m}\right),		\label{eq:Xlm}
	\end{align}
	and $Y_{l m}$ are the standard spherical harmonics, with normalization:
	\begin{equation}
		\int \int \mathrm{d} \theta \mathrm{d} \phi ~Y_{l m}^{*}(\theta, \phi) Y_{l^{\prime} m^{\prime}}(\theta, \phi)=\delta_{l l^{\prime}} \delta_{m m^{\prime}}, \label{eq:Ynorm}
	\end{equation}
	where $^*$ on the spherical harmonics denotes complex conjugation. The even-sector metric perturbation then contains 7 degrees of freedom, given by $h_0^{l m}(r)$, $h_1^{l m}(r)$, $H_0^{l m}(r)$, $H_1^{l m}(r)$, $H_2^{l m}(r)$, $K^{l m}(r)$ and $G^{l m}(r)$, where $h_0^{l m}(r)$ and $h_1^{l m}(r)$ in the even-sector are distinct to those in the odd-sector.
	As in the odd-sector, we suppress the explicit dependence on $r$ of these fields henceforth, writing the full even-sector perturbation as \cite{Berndtson:2007gsc}:
	\begin{align}
	&h_{\mu \nu}^{e, l m}=\label{evenperturbation} \\\nonumber& 
	\left(\hspace{-0.15cm}\begin{array}{cccc}{f(r)H_{0}^{l m} Y_{l m}} &{H_{1}^{l m}Y_{l m}} & {h_{0}^{l m}\frac{\partial Y_{l m}}{\partial \theta}} & {h_{0}^{l m}\frac{\partial Y_{l m}}{\partial \phi}} \\
	{*} & {\frac{H_{2}^{m}Y_{l m}}{f(r)}} & {h_{1}^{l m}\frac{\partial Y_{l m}}{\partial \theta}} & {h_{1}^{l m} X_{l m}} \\
	{*} & {*} & {r^{2}\left(K^{l m}Y_{l m}\right. } & {r^{2} \sin \theta G^{l m}X_{l m}} \\
	{} & {} & {\left.+G^{l m}W_{l m}\right)} & {} \\
	{*}&{*}&{*}&{r^{2} \sin ^{2} \theta\left(K^{l m}Y_{l m}\right.}\\
	{} & {} & {} & {\left.-G^{l m}W_{l m}\right)} 
	\end{array}\hspace{-0.25cm}\right).
	\end{align}
	The RW gauge is defined by setting the field $h_2^{l m}$ from the odd-sector and the fields $h_0^{l m}$, $h_1^{l m}$ and $G^{l m}$ from the even-sector for all $l$ and $m$ to zero \cite{Berndtson:2007gsc,PhysRev.108.1063}. 
	The $s = 2$ RW (Zerilli) equations are derived from applying the RW gauge to the odd (even)-sector of the generic first-order linearised Einstein field equations \cite{Maggiore}:
	\begin{equation}
		\delta G [h^1_{\alpha \beta}] = -16 \pi T^1_{\alpha \beta}. \label{EF1gen}
	\end{equation}
	The first-order scheme is equivalent to treating the secondary as a point particle and we solve for the retarded field \cite{Miller:2020bft, Pound:2012dk, Gralla:2012db, Pound:2014xva, Miller:2016hjv}. 
	Therefore, $T^1_{\alpha \beta}$ can be written as:
	\begin{equation}
		T_1^{\alpha \beta}(x^\mu;z^\mu) = \mu \int_{- \infty}^{\infty} \dfrac{\delta^4(x^{\mu} - z^{\mu}(\tau))}{\sqrt{-g}}\dfrac{d z^{\alpha}}{d \tau} \dfrac{d z^{\beta}}{d \tau} d \tau \label{T1}.
	\end{equation}
	The odd- and even-sector split for $T_1^{\alpha \beta}$ is given explicitly in matrix form in \Cref{tensorbasis}. The odd- and even-sector solutions for the $s = 0$, $1$ RW equations can be derived similarly from electromagnetic and scalar perturbations.
	For generic spin-weight $s$ and a given $l$, $m$ mode, the radial RWZ master equations are given by \cite{PhysRev.108.1063}: 
	\begin{equation}
		\mathcal{L}_{s}\psi_{s}(r) = S_{s}(r),\label{RWeqn2}
	\end{equation}
	where
	\begin{equation}
		\mathcal{L}_{s} \equiv \bigg(\dfrac{d^2}{dr_*^2}-V(r)+\omega^2\bigg).\label{RWeqn}
	\end{equation}
	The tortoise coordinate $r_*$ for a Schwarzschild background is defined such that $dr_*/dr = f(r)^{-1}$. Integrating this and choosing an integration constant we get: 
	\begin{align}
		r_*(r) = r + 2\log \bigg(\frac{r}{2M} - 1\bigg).
	\end{align}
	
	The RW master equation is obtained by setting the potential in \eqn{RWeqn} to:
	\begin{equation}
		V(r) = f(r)\left(\frac{l(l+1)}{r^2}+\frac{2M (1-s^2)}{r^3}\right), \label{VRW}
	\end{equation}
	where $s$ is the spin weight. Similarly, the Zerilli equation is obtained by setting the potential in \eqn{RWeqn} to:
	\begin{equation}
		V(r) = \frac{f(r)}{r^{2}\Lambda^{2}}\left[2 \lambda^{2}\left(\lambda+1+\frac{3 M}{r}\right)+\frac{18 M^{2}}{r^{2}}\left(\lambda+\frac{M}{r}\right)\right], \label{VZ}
	\end{equation}
	which is defined only for $s = 2$, and where $\Lambda = \lambda +3M/r$ and $\lambda = (l+2)(l-1)/2$.
	The RWZ master functions, $\psi_{s}(r)$ are the solutions to either the RW or Zerilli master equations and are sourced by $S_{s}(r)$. 
	The inhomogeneous RW gauge metric perturbation, $\hRW$ can then be reconstructed from the RWZ master functions \cite{PhysRev.108.1063} and \eqn{metricmodaldecomp}. 
	
	\subsection{Solving for the Regge-Wheeler and Zerilli master functions} \label{psiretsec}
 	
	For a given $l$, $m$ mode, the radial RWZ master equations, given by \eqn{RWeqn2}, can be solved using the standard method of variation of parameters.
	With this approach the inhomogeneous solution can be written as:
 	\begin{equation}
 		\psi_{s}(r) = C_s^+(r) \psi_s^{+}(r) + C_s^-(r) \psi_s^{-}(r), \label{eq:psiret2}
 	\end{equation}
 	where $\psi_s^\pm$ are the two linearly independent homogenous solutions representing ingoing and outgoing solutions radiating to asymptotic null infinity, $\mathcal{I}^+$ and the future event Horizon $\mathcal{H}^+$ respectively. Their asymptotic boundary conditions are:
	\begin{align}
		\psi_s^{+}  & \propto e^{i\omega r_*}, \hspace{0.2cm} \quad \text{as} \quad r_* \rightarrow \infty,	\label{eq:psi+_asympBC}	\\
		\psi_s^{-} & \propto e^{-i\omega r_*},  \quad \text{as} \quad r_* \rightarrow -\infty.		\label{eq:psi-_asympBC}
	\end{align}
	The weighting functions are given by:
	\begin{align}
		C^+_{s}\left(r_*\right) & = \int_{-\infty}^{r_*} \frac{\psi_{s}^{-}\left(r_*^{\prime}\right) S_{s}\left(r_*^{\prime}\right)}{W_{s}(r_*^{\prime})} \mathrm{d} r_*^{\prime}, \label{cp}\\
		C^-_{s}\left(r_*\right) & = \int_{r_*}^{\infty} \frac{\psi_{s}^{+}\left(r_*^{\prime}\right) S_{s}\left(r_*^{\prime}\right)}{W_{s}(r_*^{\prime})} \mathrm{d} r_*^{\prime}, \label{cm}
	\end{align}
	and the Wronskian, $W_s$ is defined in the usual way:
	\begin{equation}
		W_{s}(r_*) \equiv \frac{d \psi_{s}^{+}\left(r_{*}\right)}{d r_{*}} \psi_{s}^{-}\left(r_{*}\right)-\frac{d \psi_{s}^{-}\left(r_{*}\right)}{d r_{*}} \psi_{s}^{+}\left(r_{*}\right).
	\end{equation}
	As there are no first derivatives with respect to $r_*$ in \eqn{RWeqn2}, $W_s$ is a constant by Abel's theorem. 
	Accordingly, we drop the dependence on $r_*$ of $W_s$.
	
	As the secondary can be treated as a point-like particle to first-order in $\e$ \cite{Miller:2020bft, Barack:2018yvs}, the sources of the RWZ master functions take the form:
	\begin{equation}
		S_{s}(r) = p_{s}(r,\rp)\delta(r-\rp) +q_{s}(r,\rp)\delta^{\prime}(r-\rp), \label{Sdeltadeltaprime}
	\end{equation}
	where a prime denotes differentiation with respect to $r$. In the case of quasicircular orbits, at leading order we have:
	\begin{equation}
		S_{s}(r) = p_{s}(r,r_0)\delta(r-r_0) +q_{s}(r,r_0)\delta^{\prime}(r-r_0) \label{Sdeltadeltaprimecirc},
	\end{equation}
	where the factors $p_{s}(r,r_0)$ and $q_{s}(r,r_0)$ can be determined from Eqs.~\eqref{S1RW}-\eqref{Se0bRW} and Eqs.~\eqref{So02circ}-\eqref{Se22circ}. 
	For example, the leading order source for the spin-weight $s = 2$ RW master function in the odd-sector is \cite{Berndtson:2007gsc}:
	\begin{align}
		S^{lm}_{2}(r) \label{S2RWtext} = &~ \frac{4i m \pi f(r)(2 \delta(r-r_0)+r f(r)\delta'(r-r_0))}{\lambda (\lambda+1) r r_0^3}\\&\nn\times\sqrt{\frac{r_0}{r_0-3}}~\partial_\theta Y^*_{lm}\left(\frac{\pi}{2},0\right).
	\end{align}
	We can immediately read off $p_s(r,r_0)$ and $q_s(r,r_0)$ as:
	\begin{align}
	p_2(r,r_0) & = \frac{ 8 i m \pi f(r)}{\lambda (\lambda+1) r r_0^3}\sqrt{\frac{r_0}{r_0-3}}~\partial_\theta Y^*_{lm}\left(\frac{\pi}{2},0\right),\label{p}\\
	q_2(r,r_0)& = \frac{ 4 i m \pi f(r)^2}{\lambda (\lambda+1) r_0^3}\sqrt{\frac{r_0}{r_0-3}}~\partial_\theta Y^*_{lm}\left(\frac{\pi}{2},0\right).\label{q}
	\end{align}
	Due to the distributional form of $S_s$ in \eqn{Sdeltadeltaprimecirc}, \eqn{cp} and \eqn{cm} become:
	\begin{align}
		C_s^+(r) &= c_s^+\Theta(r - r_0), \label{Cp}\\ 
		C_s^-(r) &= c_s^-\Theta(r_0 - r), \label{Cm}
	\end{align}
	where integration by parts has been used to evaluate terms in \eqn{cp} and \eqn{cm} involving $\delta^{\prime}$, and $\Theta$ is the Heaviside step function, defined so that $\Theta(x) = 1$ for $x> 0$ and $\Theta(x) = 0$ for $x \leq 0$. The constants $c_s^\pm$ are the weighting coefficients: 
	\begin{align}
	\label{c} c_s^{\pm} =~ \frac{1}{W_s} \bigg\{&\frac{\psi_{s}^{\mathrm{\mp}}\left(r_0\right)  p_s(r,r_0)}{f(r_0)}\\&\nonumber-\frac{\partial}{\partial r}\left(\frac{\psi_{s}^{\mathrm{\mp}}\left(r\right) q_s(r,r_0)}{f(r)}\right) \bigg\}\Bigg\rvert_{r=r_0},
	\end{align}
	and the radial derivatives of $\psi_s$ can be determined by taking the radial derivative of \eqn{eq:psiret2}. 
	
	\subsection{Calculating the $r_0$ derivative of the Regge-Wheeler and Zerilli master functions} \label{psirretsec}
	
	For quasicircular orbits, during the inspiral, slow-time derivatives are equivalent to derivatives with respect to $r_0$, with an additional factor of $dr_0/d\tilde{t}$. 
	We wish to solve for the slowly evolving RWZ master functions, $\partial_{r_0} \psi_s$, described by the following equation, obtained by taking the $r_0$ derivative of \eqn{RWeqn2}:
	\begin{equation}
		\mathcal{L}_s\dpsi_s = S_{s,r_0} - 2 \omega \omega_{,r_0} \psi_s, \label{noncompact1}
	\end{equation}
	where $\dpsi_s \equiv \psi_{s,r_0}$ and a comma followed by $r_0$ denotes a derivative with respect to $r_0$. 
	Unlike $S_s$, the source in \eqn{noncompact1} is no longer compact, as $\psi_s$ is defined over the entire domain. 
	There are a variety of numerical techniques for finding solutions to equations of the form of \eqn{noncompact1} e.g., \cite{Berndtson:2007gsc,Miller_etal,PanossoMacedo:2022fdi}, but none of them are as efficient or easy to implement as solving an equation with a distributional source. 
	We can find an equation for $\dpsi_s$ with a distributional source by noticing that applying the operator $\mathcal{L}_s$ to the second term on the right hand side of \eqn{noncompact1}, then making use of \eqn{RWeqn2}, compactifies that term.
	Applying an additional operator $\mathcal{L}_s$ to \eqn{noncompact1} therefore `partially annihilates' the non-compact term on the right hand side \cite{Hopper:2012ty, Mathews:2021rod}, yielding a fourth-order differential equation with a compact, distributional source: 
	\begin{equation}
		\mathcal{L}_s^2\dpsi_{s} = \mathcal{L}_s S_{s,r_0} - 2 \omega \omega_{,r_0} S_s. \label{slowtimeRW}
	\end{equation}
	Thus we obtain a family of fourth-order ODEs with compact sources, which we shall solve using the method of variation of parameters. 
	As a fourth-order differential equation, there are four independent homogeneous solutions. 
	Two of these are the homogeneous solutions to Eq.~\eqref{RWeqn2} as $\mathcal{L}_s(\mathcal{L}_s(\psi_s^\pm)) = \mathcal{L}_s(0) = 0$.
	The remaining two we denote by $\dpsi^{h4,\pm}_s$.
	These are homogeneous solutions to the fourth-order equation Eq.~\eqref{slowtimeRW} but not the second-order equation, Eq.~\eqref{RWeqn2}.	
	The homogeneous solutions have the asymptotic boundary conditions:
	\begin{align}
		\dpsi_s^{h4,+}  & \propto r e^{i\omega r_*},	\quad  \text{as} \quad r_* \rightarrow \infty,			\label{eq:dpsi+_BC}\\
		\dpsi_s^{h4,-} & \propto \log(f(r)) e^{-i\omega r_*},	\quad \text{as} \quad r_* \rightarrow -\infty. \label{eq:dpsi-_BC}
	\end{align}
	
	Similarly to \eqn{eq:psiret2}, and due to the distributional source, we can write the inhomogeneous solution as:
	\begin{align}
		\dpsi_{s}(r) = & \left[c_s^{h2,+} \psi_{s}^{+}(r) + c_s^{h4,+} \dpsi_{s}^{h4,+}(r)\right]\Theta(r - r_0) \nonumber\\ \nonumber
		& +\left[c_s^{h2,-} \psi_{s}^{-}(r) + c_s^{h4,-} \dpsi_{s}^{h4,-}(r)\right]\Theta(r_0 - r) \\
		& + c^\delta_s \dpsi_{s}^{\delta}(r)\delta(r - r_0), \label{psirret}
	\end{align}
	where the constant coefficients (for a given $s,l,m,r_0$) are given by:
	\begin{align}
		c_s^{i,+}\Theta(r - r_0) &=   \int_{-\infty}^{r_*} \hspace{-0.1cm}\frac{\mathcal{W}_{s}^{i,+}\left(r_*^{\prime}\right) \mathcal{S}_{s}\left(r_*^{\prime}\right)}{\mathcal{W}_{s}} \mathrm{d} r_*^{\prime},  \label{eq:ch4+}\\
		c_s^{i,-}\Theta(r_0 - r) &=   \int_{r_*}^{\infty} \frac{\mathcal{W}_{s}^{i,-}\left(r_*^{\prime}\right) \mathcal{S}_{s}\left(r_*^{\prime}\right)}{\mathcal{W}_{s}} \mathrm{d} r_*^{\prime},	\label{eq:ch4-}
	\end{align}
	where $i \in \{h2,h4\}$ and we have defined:
	\begin{equation}
		\mathcal{S}_{s} \equiv \mathcal{L}_s S_{s,r_0} - 2 \omega \omega_{,r_0} S_s. \label{stsource}
	\end{equation} The fourth-order Wronskian $\mathcal{W}_s$ is also constant by Abel's Theorem and has the standard definition:
	\begin{equation}
		\hspace{-0.2cm} \mathcal{W}_s = \det\left(\begin{array}{cccc}
		{\psi_s^{-}}&{\phi_s^{h4, -}}&{\psi_s^{+}}&{\phi_s^{h4, +}}\\
		{\partial_{r_*}\psi_s^{-}}&{\partial_{r_*}\phi_s^{h4,-}}&{\partial_{r_*}\psi_s^{+}}&{\partial_{r_*}\phi_s^{h4,+}}\\
		{\partial^2_{r_*}\psi_s^{-}}&{\partial^2_{r_*}\phi_s^{h4,-}}&{\partial^2_{r_*}\psi_s^{+}}&{\partial^2_{r_*}\phi_s^{h4,+}}\\
		{\partial^3_{r_*}\psi_s^{-}}&{\partial^3_{r_*}\phi_s^{h4,-}}&{\partial^3_{r_*}\psi_s^{+}}&{\partial^3_{r_*}\phi_s^{h4,+}}
		\end{array}\right),\label{Ws4}
	\end{equation}
	and $\mathcal{W}_s^{h2/4,\pm}$ are given by the determinant of the matrices obtained from deleting the final row and the column containing the corresponding field, $\psi_s^{h2/4,\pm}$ and its derivatives from the matrix in \eqn{Ws4} \cite{ODEs,Hopper:2012ty}, where we have defined $\psi^{h2,\pm}_s \equiv \psi^\pm_s$.
	The field $\dpsi_{s}^{\delta}(r)$ in Eq.~\eqref{psirret} is required to balance the forth derivative of the delta function that appears in the source in \eqn{slowtimeRW}.
	
	Owing to the distributional source, the integrals in Eqs.~\eqref{eq:ch4+} and \eqref{eq:ch4-} can be carried out analytically by repeated application of integration by parts.
	An equivalent approach is to match the coefficients of the delta function and its derivatives between the left- and right-hand side of Eq.~\eqref{slowtimeRW}.
	Following from \eqn{Sdeltadeltaprimecirc} and \eqn{stsource}, $\mathcal{S}_s$ takes the form:
	\begin{align}
		\mathcal{S}_s  = & ~a_0(r,r_0) \delta(r-r_0) + a_1(r,r_0) \delta^{\prime}(r-r_0) \label{RHS1}\\&\nonumber+a_2(r,r_0) \delta^{\prime \prime}(r-r_0) + a_3(r,r_0) \delta^{\prime \prime \prime}(r-r_0) \\&\nonumber + a_4(r,r_0) \delta^{(4)}(r-r_0),
	\end{align}
	where $a_0(r,r_0), \dots , a_1(r,r_0)$ can be identified with combinations of $p(r,r_0)$, $q(r,r_0)$ and their derivatives with respect to $r$ and $r_0$ by replacing the $S_s$ in \eqn{stsource} with the right-hand side of \eqn{Sdeltadeltaprimecirc}. Making use of the following identity:
	\begin{equation}
	f(r) \delta(r - r_0) = f(r_0) \delta(r - r_0) \label{deltaeqn},
	\end{equation}
	and other identities derived from radial derivatives of \eqn{deltaeqn}, the source $\mathcal{S}_s$ can be written in terms of Dirac delta functions and their radial derivatives with constant coefficients:
	\begin{align}
		\mathcal{S}_s  = & ~b_0(r_0) \delta(r-r_0) + b_1(r_0) \delta^{\prime}(r-r_0) \label{RHS2}\\&\nonumber+b_2(r_0) \delta^{\prime \prime}(r-r_0) + b_3(r_0) \delta^{\prime \prime \prime}(r-r_0) \\&\nonumber  + b_4(r_0) \delta^{(4)}(r-r_0), 
	\end{align}
	where $b_0(r_0), \dots , b_4(r_0)$ can be identified with combinations of $a_0(r_0), \dots , a_4(r_0)$ and their radial derivatives, evaluated at $r = r_0$ such that they are now functions of $r_0$ only. 
	As an example, the coefficients of the Dirac delta functions and their derivatives are given explicitly for the odd-sector source for spin-weight $s =2$ in \Cref{b's}. 
	Substituting \eqn{psirret} into \eqn{slowtimeRW} and making use of \eqn{deltaeqn} we obtain:
	\begin{align}
		\mathcal{L}_s^2 \dpsi_{s}  = & ~ \beta_1(r) \Theta(r-r_0) + \beta_2(r) \Theta(r_0-r) \label{LHS1} \\&
		\nonumber+c_0(r_0) \delta(r-r_0) + c_1(r_0)\delta^{\prime}(r-r_0)\\&
		\nonumber +c_2(r_0) \delta^{\prime \prime}(r-r_0) + c_3(r_0) \delta^{\prime \prime \prime}(r-r_0) \\&\nonumber  + c_4(r_0) \delta^{(4)}(r-r_0),
	\end{align}
	where $c_0(r_0), \dots , c_4(r_0)$ can be identified with combinations of $c_s^{h2/4, \pm}$ and $c_s^\delta$. 
	Immediately we find $\beta_1(r) = \beta_2(r) = 0$ as these coefficients satisfy the homogeneous field equation, and there are no Heaviside terms on the right-hand side of \eqn{slowtimeRW}. 
	Equating the constant coefficients of the Dirac delta functions and their radial derivatives from \eqn{LHS1} with those in \eqn{RHS2}, we obtain a linear system which we can solve for $c_s^{h2/4, \pm}$ and $c_s^\delta$ in terms of $p(r_0)$, $q(r_0)$ and their derivatives, evaluated at $r = r_0$. We can write this linear system of equations in the following form:
	\begin{align}
		\mathbf{C}_s = \mathbf{\Phi}_s^{-1}\cdot\mathbf{J}_s,	\label{eq:CsFromJumps}
	\end{align}
where $\mathbf{C}_s =(c_s^{h2,-},c_s^{h4,-},c_s^{h2,+},c_s^{h4,+})$ and
\begin{equation}
	\mathbf{\Phi}_s=\left(\begin{array}{cccc}
	{-\psi_s^{-}}				&{-\dpsi_s^{h4, -}}				&{\psi_s^{+}}				&{\dpsi_s^{h4, +}}\\
	{-\partial_{r}\psi_s^{-}}	&{-\partial_{r}\dpsi_s^{h4,-}}	&{\partial_{r}\psi_s^{+}}	&{\partial_{r}\dpsi_s^{h4,+}}\\
	{-\partial^2_{r}\psi_s^{-}}	&{-\partial^2_{r}\dpsi_s^{h4,-}}	&{\partial^2_{r}\psi_s^{+}}	&{\partial^2_{r}\dpsi_s^{h4,+}}\\
	{-\partial^3_{r}\psi_s^{-}}	&{-\partial^3_{r}\dpsi_s^{h4,-}}	&{\partial^3_{r}\psi_s^{+}}	&{\partial^3_{r}\dpsi_s^{h4,+}}
	\end{array}\right)_{r=r_0}.
	\end{equation}
	The vector $\mathbf{J}_s$ is given by $\mathbf{J} = (J_0, J_1, J_2, J_3)$ where $J_n$ is the jump in the $n^\text{th}$ derivative of $\dpsi_{s}$ on the world-line.
	We give the explicit form of $\dpsi_s^\delta(r_0)$ and these jumps for the $s=2$ RW field in odd-sector in \Cref{apdx:Js}.
	The jump conditions for the Zerilli master function are too long to appear in this article to we have included them in the supplemental material \cite{suppmat}.

	\subsection{Numerical boundary conditions and implementation}\label{sec:RWZ_BCs}

	When solving for perturbations on hypersurfaces of constant $t$ we cannot place numerical boundary conditions at future null infinity, $\mathcal{I}^+$, and the future event Horizon $\mathcal{H}^+$.
	Instead we will find series expansions of the solutions at finite radii.
	For the homogeneous solutions to the RWZ equations we expand the asymptotic boundary conditions in Eqs.~\eqref{eq:psi+_asympBC} and \eqref{eq:psi-_asympBC} as:
	\begin{align}
		\hspace{-0.2cm}\psi_{s}^{+}(r) =& e^{i \omega r_{*} }\hspace{-0.1cm} \sum _{i=0}^{n_{\text{max}}} \frac{A^{+}_i}{(r \omega)^{i}}\bigg\vert_{r = r_{\text{out}}},		\label{eq:psi+_numBC}\\
		\hspace{-0.2cm}\psi_{s}^{-}(r) =& e^{-i \omega r_*} \hspace{-0.2cm} \sum _{i=0}^{n_{\text{max}}}A^{-}_if(r)^i\big\vert_{r = r_{\text{in}}}, 						\label{eq:psi-_numBC}
	\end{align}
	where the coefficients $A^{\pm}_i$ depend on the parameters $s$, $l$, $m$, $r_0$. In order for the expansion in Eq.~\eqref{eq:psi+_numBC} to converge, $r_{\text{out}}$ must be in the wave zone such that $r \omega \gg 1$ \cite{poisson_will_2014}.
	In practice we find that $r_{\text{out}} = 10^4M$ is sufficient for the parameters we consider in this work.
	Similarly we find that $r_{\text{in}} = (2 + 10^{-5})M$ ensures rapid convergence of the expansion in \eqn{eq:psi-_numBC}.
	We shall use these values of $r_\text{in/out}$ though out this work.
	We also find a value of $n_\text{max} = 50$ is sufficient to ensure the series expansions satisfy the homogeneous field equation to beyond machine precision.
	We find this value of $n_\text{max}$ is sufficient for all later boundary condition expansion as well.
	By substituting the above expansions into the homogeneous field equation, \eqn{RWeqn2}, recurrence relations can be derived for the coefficients $A^\pm_i$ in terms of the leading $i=0$ coefficient.
	In our code we use the \texttt{ReggeWheeler} package of the BHPToolkit \cite{BHPToolkit} to compute $\psi^{\pm}_s$.
	The numerical integration option in this package implements the above boundary condition expansions and then numerically integrates to $r_0$ to find the homogeneous solutions at any radius.
	We note here that the \texttt{ReggeWheeler} package uses the Mano-Suzuki-Takasugi (MST) method by default, which computes $\psi_s^\pm$ as a series of hypergeometric functions \cite{Mano:1996vt, Casals:2015nja}.
	This method allows for very high precision numerical results to be obtained but has the downside that it often requires extended precision arithmetic and typically is slower to compute the homogeneous solution at a given radius for strong-field orbits.
	As the goal of this work is to provide $\hLr$ on a dense grid of $r$ values for use in constructing the source to second-order perturbations, we instead opt to use the faster numerical integration method.
	
	The asymptotic boundary conditions for $\dpsi_s^{h4,\pm}$, given in Eqs.~\eqref{eq:dpsi+_BC} and \eqref{eq:dpsi-_BC}, can be expanded as:
	\begin{align}
		\hspace{-0.2cm}\dpsi_{s}^{h4,+}(r) =& r e^{i \omega r_{*} }\hspace{-0.1cm} \sum _{i=0}^{n_{\text{max}}} \frac{[A^{h4,+}_i\hspace{-0.1cm}+B^{h4,+}_i\log(r)]}{(r \omega)^{i}}\bigg\vert_{r = r_{\text{out}}},\label{eq:RWout}\\
		\hspace{-0.2cm}\dpsi_{s}^{h4,-}(r) =& e^{-i \omega r_*} \hspace{-0.2cm} \sum _{i=0}^{n_{\text{max}}}[A^{h4,-}_i\hspace{-0.1cm}+B^{h4,-}_i\log(f(r))]\label{eq:RWin}f(r)^i\big\vert_{r = r_{\text{in}}}.
	\end{align}
	As before, the coefficients $A^{h4, \pm}_i$ and $ B^{h4, \pm}_i$ depend on the parameters $s$, $l$, $m$, $r_0$ and are derived by substituting the above ans\"atze into the fourth-order equations \eqn{slowtimeRW}.
	The recurrence relations for these coefficients is provided in the supplementary material \cite{suppmat}, and depend on the choice of potential, either RW or Zerilli. 
	The series approximation to the homogeneous solutions near the boundaries are then given in terms of one of the leading coefficients.
	For the expansion at large radius for both the RW and Zerilli cases we find $A_0^{h4,+} = B_1^{h4,+}/2$ and $B_0^{h4,+} = 0$.
	For the series approximation that satisfies the fourth-order equation, but not the second-order equation, we find we can set, e.g., $B_1^{h4,+} = 1$ and $ A_1^{h4,+} = 0$.
	Note if we set $B_1^{h4,+} = 0$ and $A_1^{h4,+} = 1$ we recover the boundary condition expansion in Eq.~\eqref{eq:psi+_numBC} above for the second-order equation.
	For the expansion near the horizon we find that setting $A^{h4,-}_0 = 0$ and $B^{h4,-}_0 = 1$ provides an approximate solution to the fourth-order equation (but not the second-order equation).
	Setting $A^{h4,-}_0 = 1$ and $B^{h4,-}_0 = 0$ we similarly recover the boundary condition in Eq.~\eqref{eq:psi-_numBC}.
	Once the value of the leading terms is set, all other coefficients are then determined by the recurrence relations.
	These homogeneous solutions are not calculated by any package in the BHPToolkit so we solve the recurrence relations with the above conditions.
	This then provides the boundary conditions at $r_{\text{out}}$ and $r_{\text{in}}$ and we use the \texttt{NDSolve} function of \textit{Mathematica} to numerically integrate the solutions to $r_0$.

		 \subsection{The $r_0$ derivative of the flux to infinity}\label{sec:RW_r0deriv_Flux}
	 	 
	In order to calculate the GW energy flux radiated to null infinity we consider perturbations with constant retarded time, $u=t-r_*$.
	These perturbations can be related to the perturbations on constant $t$ slices via:  
	\begin{align}
		\psi_2^{lm,[u]} = \psi_2^{lm} e^{-i\omega r_*}.	\label{eq:psi_u}
	\end{align}
	The GW energy flux radiated to infinity can be calculated via \cite{Barack:2010tm,PhysRevD.82.084010}:
	\begin{align}
		\dot E^{l m}_\infty &= \frac{\lambda l(l+1)}{8 \pi} g(r_0) |\psi_{2}^{lm,[u]}|_{r\rightarrow \infty}^2, \label{eq:flux}\\
	 					    &= \frac{\lambda l(l+1)}{8 \pi} g(r_0) |c^{+}_{2,lm}|^2, 
	\end{align}
	where $g(r_0) = 1$ for the RW case in the odd-sector, $g(r_0) = (\omega/2)^2$ for the Zerilli case in the even-sector, and the limit to infinity is taken with respect to fixed retarded time.
	Taking an $r_0$ derivative of \eqn{eq:flux} we obtain:
	\begin{align}
		\partial_{r_0} \dot E^{l m}_\infty =& \frac{\lambda l(l+1)}{8\pi} \bigg\{2g(r_0)\text{Re}\left[\dpsi_{2}^{lm,[u]} \psi^{lm,[u]*}_2\right]		\nonumber \\
		& + g'(r_0) |\psi_2^{lm,[u]}|^2\bigg\}_{r \rightarrow \infty}.\label{dr0flux}
	\end{align}
	To evaluate this formula we start by taking the $r_0$ derivative of Eq.~\eqref{eq:psi_u} we get:
	\begin{align}
		\dpsi_{2}^{lm,[u]} = e^{-i \omega r_*}\left(\dpsi_{2}^{lm} -i r_* \omega_{,r_0}\psi_2^{lm}\right).
	\end{align}
	Now using Eqs.~\eqref{eq:psiret2}, \eqref{psirret} and the following leading terms in the boundary condition expansions  Eqs.~\eqref{eq:psi+_numBC} and \eqref{eq:RWout}, given by:
	\begin{align}
		\psi_2^{lm} e^{-i \omega r_*}		&\sim c_2^+\left[1 + \frac{i l(l+1) r_0^{3/2}}{2 m r} \right], 		\label{eq:psi2u-leadingBC}\\
		\dpsi_{2}^{lm} e^{-i \omega r_*}	&\sim c_2^{h2,+} + c^{h4,+}_2 \left[\frac{r}{2} + \log(r)\right],	\label{eq:phiExp-leadingBC}
	\end{align}
	we get:
	\begin{align}
		\dpsi_{2}^{lm,[u]} \sim& \left(c^{h4,+}_2 + \frac{3 i c_2^+ m}{r_0^{5/2}}\right)\left[\frac{r}{2} + \log(r)\right] + c_2^{h2,+} \nonumber \\ 
								  & - c_2^+ \left[\frac{3 l(l+1)}{4 r_0} + \frac{3i m \log(2)}{r_0^{5/2}}\right].
	\end{align}
	Note that the expansions given in Eqs.~\eqref{eq:psi2u-leadingBC} and \eqref{eq:phiExp-leadingBC} are valid for both the RW and Zerilli potentials.
	In order for this result to be finite, when we take the limit to infinity we must have:
	\begin{align}
		c^{h4,+}_2 = -\frac{3 i c_2^+ m}{r_0^{5/2}}.	\label{eq:ch4c2relation}
	\end{align}
	Combining the above results, the GW flux to infinity in the odd-sector is given by:
	\begin{align}
	\label{eq:dr0Edot_RW} \partial_{r_0} \dot E^{lm,RW}_\infty =&  \frac{\lambda l(l+1)}{4\pi} \bigg[\text{Re}\left(c_2^{h2,+}c_2^{+*}\right)
		\\ &\nonumber- |c_2^+|^2 \left(\frac{3l(l+1)}{4 r_0}\right)\bigg],
	\end{align}
	and in the even-sector by:
	\begin{align}
		\partial_{r_0} \dot{E}_\infty^{lm,Z} = \frac{m^2}{4 r_0^3}\left(\partial_{r_0} \dot E^{lm,RW}_\infty  - \frac{6 |c_2^+|^2}{4r_0}\right). \label{eq:dr0Edot_Z}
	\end{align}
	
	\subsection{Numerical Results}\label{sec:RW_results}
	
\begin{figure}[h]
	\includegraphics[width=0.48\textwidth]{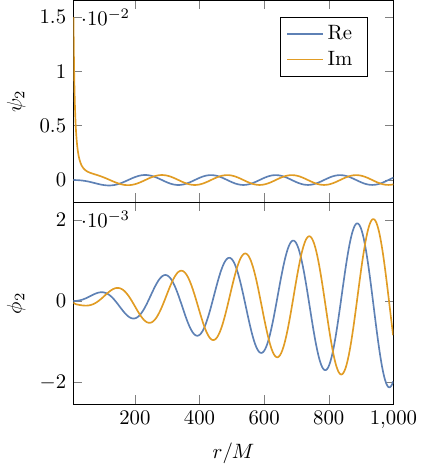}
	\caption{
		Real and imaginary parts of $\psi_2$ (top panel) and $\phi_2 = \psi_{2,r_0}$ (bottom panel) for the $(l,m)=(2,1)$ mode with $r_0 = 10M$. For the latter case, the amplitude of the wave grows proportional to $r$ for large $r$, as determined by the asymptotic boundary condition in \eqn{eq:RWout}.
		} \label{Reim}
	\end{figure}

	Using the homogeneous solutions computed numerically as described in Sec.~\ref{sec:RWZ_BCs} and the jump conditions from Eq.~\eqref{eq:CsFromJumps}, \Cref{apdx:Js} and the supplementary material \cite{suppmat}, we obtained numerical results for the $r_0$ derivatives of all of the RWZ master functions.
	As an example, in \Cref{Reim} we plot the results of the spin-weight $s = 2$ RW master function for the $(l,m) = (2,1)$ mode at $r_0 = 10 M$ out to large $r$. 
	
\begin{figure}[h]
	\includegraphics[width=0.48\textwidth]{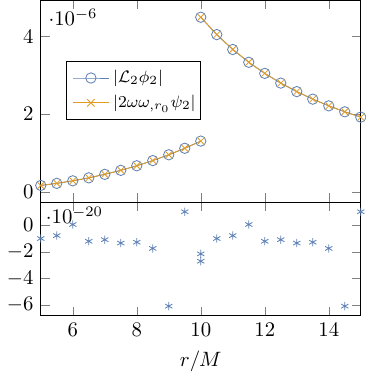}
	\caption{Absolute values of the left and right hand sides of \eqn{noncompact1}. The top panel demonstrates that our calculation of $\phi_2$ (where $\phi_2 = \psi_{2,r_0}$) for $r_0 = 10 M$ and $(l,m) = (2,1)$ solves \eqn{noncompact1}, not including terms involving the Dirac delta functions and its derivatives. The lower panel shows the absolute difference between the data sets for the left- and right-hand side of \eqn{noncompact1}. This error is dominated by the interpolation order of the numerical homogenous solutions and their radial derivatives.} \label{dr0psi2RWcombined}
	\end{figure}

	To test the results of our partial annihilator method, we check that $\dpsi_{2}$ satisfies the original second-order equation with a non-compact source, \eqn{noncompact1}. 
	Our numerical integrator returns the $\dpsi_2$ and its first, second, and third radial derivatives which we use to compute the left-hand side of \eqn{noncompact1}.
	We compute the right-hand side of \eqn{noncompact1} using the \texttt{ReggeWheeler} package of the BHPToolkit \cite{BHPToolkit} and find this matches the left-hand side to near machine precision.
	We give an example of this for the $(l,m)=(2,1)$ mode in \Cref{dr0psi2RWcombined}.
	
 As a further check that our results are working correctly, we compute $\partial_{r_0}\dot{E}^{lm}_\infty$ using \eqn{eq:dr0Edot_RW} and \eqn{eq:dr0Edot_Z} and compare these results to the $r_0$ derivative of the flux when computed numerically.
 To compute the latter we compute the flux on a dense grid of $r_0$ values using the \texttt{ReggeWheeler} package \cite{BHPToolkit}.
 Using the standard least-squares algorithm we fit this data to an eighth-order polynomial centred on the $r_0$ value where we wish to compute the $r_0$ derivative of the flux.
 The linear coefficient in this fit is the numerical approximation to the $r_0$ derivative of flux.
 We find excellent agreement between these two methods and present some sample numerical results in Table~\ref{table:dr0Edot}.
 We also numerically check that the relation in Eq.~\eqref{eq:ch4c2relation} between $c_2^{h4,+}$ and $c_2^{+}$ holds to machine precision.
 
 To assess the efficiency of our partial annihilator method while avoiding issues regarding the level of code optimisation between our code the BHPToolkit we can count the number of times our approach must solve an ODE verses the number of times required when computing the $r_0$ derivative numerically. Numerically computing the homogeneous basis of solutions is by far the largest computational cost as all the other steps, such as matching to get the inhomogeneous solutions or calculating the flux, are quick algebraic operations. Computing $\psi_2$ using the partial annihilator method requires solving the left-hand side of Eq.~\eqref{slowtimeRW} four times. To numerically compute the $r_0$ derivative of the flux at, e.g., $r_0 = 8M$, to a relative precision of $\sim10^{-12}$ we find we must compute the flux at 11 equally spaced radii between in the range $r_0 \pm 0.2M$, and fit to an eighth-order polynomial. This requires solving the left-hand side of Eq.~\eqref{RWeqn} 22 times. Thus our method is $22/4 =5.5$ faster to compute $\partial_{r_0}\dot{E}^{lm}_\infty$ than the numerical derivative approach.  In practice, when performing the Regge-Wheeler to Lorenz transformation the speed up is closer to a factor of 10 as two of the homogeneous solutions to Eq.~\eqref{slowtimeRW} are $\psi_2^\pm$ which will already be computed for other parts of the transformation. We also note that in order to reach a relative precision beyond $\sim10^{-9}$ in the numerical $r_0$ derivative computed using the \texttt{ReggeWheeler} BHPToolkit package we had to use extended precision arithmetic.
 
 \begin{table}[]
 \def\arraystretch{1.3}
 \begin{tabular}{c|c|c}
  $r_0/M$& $\frac{M^3}{\mu^2}\partial_{r_0}\dot{E}_\infty$  & rel.~err.   \\
  \hline
  6 	&  	$-8.859960015046\times10^{-4}$	&  $4.1\times 10^{-11}$  \\
  8 	&  	$-1.291224908187\times10^{-4}$	&  $1.2\times 10^{-12}$  \\
  10 	&  	$-3.155702796251\times10^{-5}$	&  $1.2\times 10^{-13} $ 
 \end{tabular}
 \caption{Numerical results for the $r_0$ derivative of the GW energy flux to null infinity, with modes summed up to $l_\text{max} = 15$.
 The first column gives the orbital radius, the second column shows the $r_0$ derivative of the flux radiated to infinity as computed from our partial annihilator calculation.
 The third column shows the relative error compared to a result computed by computing the flux at many orbital radii and taking a numerical derivative as described in the main text.}\label{table:dr0Edot}
 \end{table}
	
	\section{Transforming to the Lorenz gauge}\label{parital annihilators}
	
	In this section we will review the Lorenz gauge, defined at leading order by the gauge condition:
	\begin{equation}
		\bar{h}_{\alpha \beta}^{1 ~;\beta}=0, \label{Lorenzcondition}
	\end{equation}
	where the trace-reversed metric is defined via:
	\begin{equation}
		\bar{h}_{\alpha \beta}^{n} \equiv h_{\alpha \beta}^{n}-\frac{1}{2} g^0_{\alpha \beta} g_0^{\mu \nu} h_{\mu \nu}^{n}. \label{eq:tracereversedmetric}
	\end{equation}
	We shall assume the trace-reversed metric refers to Lorenz gauge perturbations throughout this paper. 
	The first-order Einstein field equations can then be written in the Lorenz gauge as \cite{Barack:2005nr}:	 
	\begin{equation}
		\Box \bar{h}^1_{\alpha \beta} + 2 \tensor{R}{^{\mu}_{ \alpha}^{ \nu}_{\beta}} \bar{h}^1_{\mu \nu}=16 \pi T^1_{\alpha \beta}, \label{EFL}
	\end{equation}
	where $\Box\equiv\nabla_\alpha\nabla^\alpha$ is the covariant D'Alambertian operator, $\tensor{R}{^{\alpha \beta \mu \nu}}$ is the Riemann tensor for Schwarzschild spacetime, and $T^1_{\alpha \beta}$ is the stress-energy tensor of a point particle given in Eq.~\eqref{T1}. There are a variety of ways the Lorenz gauge metric perturbation has been computed. In Schwarzschild spacetime, decomposing the metric perturbation onto a basis of tensor spherical harmonics decouples the angular behaviour. The resulting set of 1+1 partial differential equations has been solved for circular \cite{Barack:2007tm} and eccentric \cite{Barack:2010tm} orbital motion. A decomposition into azimuthal $m$-modes has allowed the perturbation to be computed for circular orbits in the Schwarzschild \cite{Dolan:2012jg} and Kerr \cite{Isoyama:2014mja} spacetimes. By expanding the perturbation in the form of Eq.~\eqref{eq:LorenzFourierDecomp}, the Lorenz gauge metric perturbation has also been computed in the frequency domain for circular \cite{Akcay:2010dx} and eccentric orbits \cite{Akcay:2013wfa,Osburn:2014hoa}. In this case the radial behaviour decouples and leaves a set of coupled ODEs of the form of Eq.~\eqref{eq:LorenzGaugeRadial}. An alternative approach, that we shall pursue in this work, is to calculate the perturbation in a different gauge and transform to the Lorenz gauge \cite{Berndtson:2007gsc, Dolan:2021ijg, Hopper:2012ty,PhysRevD.69.084019}
	
	One approach to compute $\hLr$ is to take the $r_0$ derivative of \eqn{eq:LorenzFourierDecomp} to get \eqn{eq:r0-deriv_Lorenz}.
	As discussed in the introduction, \eqn{eq:r0-deriv_Lorenz} is challenging to solve due to the unbounded source and the coupling between the different components of the metric perturbation.
	This approach has been pursued in Ref.~\cite{Miller_etal}.
	In this work we shall instead use the gauge transformation approach.
	The transformation from the RWZ metric perturbation to the Lorenz gauge was first written down explicitly by Berndtson \cite{Berndtson:2007gsc}.
	More recently, the homogeneous case of these transformations were derived from an approach to compute the Lorenz gauge perturbation using the Teukolsky formalism \cite{Dolan:2021ijg}.
	In this work we will compute $\hLr$ by taking the $r_0$ derivative of Berndtson's gauge transformation. 
	We will calculate $\hL$ and $\hLr$ for quasicircular orbits in terms of the RWZ master functions $\psi_{s}$, their $r_0$ derivatives $\dpsi_s\equiv\psi_{s,r_0}$, and their respective radial derivatives, whose solutions are obtained from \Cref{RW}.
	
	In the odd-sector, the fields $\psi_1,\psi_2$ and their radial derivatives are sufficient to compute $\hL$ \cite{Berndtson:2007gsc}.
	With the addition of $\dpsi_1,\dpsi_2$ it is also possible to compute $\hLr$ as described in Sec.~\ref{Lorst} below.
	In the even-sector we must solve for the fields $\psi_{2\text{Z}}$, $\psi_1$, $\psi_0$, $\psi_{0b}$ and $\mtaf$ \cite{Berndtson:2007gsc}. The field $\psi_{2Z}$ refers to the Zerilli field, defined for $s = 2$ only, with all remaining fields in the even-sector subject to the RW potential. The field $\psi_{0b}$ obeys the same field equation as $\psi_0$ but with a different distributional source \cite{Berndtson:2007gsc}, given in \Cref{Sources}.
	The additional field, $\mtaf$ (following Berndton's notation) arises from the second term on the right-hand side of \eqn{gaugetransformeqn}, and hence is a `pure gauge' contribution to the Lorenz gauge metric perturbation.
	Unlike all the other fields, $\mtaf$ has a source which is unbounded.
	Fortunately, the source for $\mtaf$ is such that we can again apply the method of partial annihilators as we describe in Sec.~\ref{mtafsec} below. Taking an $r_0$ derivative of Berntdson's expressions for $\hL$ in the even-sector then allows us to construct $\hLr$ from the above fields, in addition to $\dpsi_{2\text{Z}}$, $\dpsi_1$, $\dpsi_0$, $\dpsi_{0b}$ and $\mtafr$. We compute $\mtafr$ using the method of partial annihilators to arrive at two sixth-order ODEs with distributional sources, described in Sec.~\ref{mtafrsec} below.

	\subsection{Calculating the Lorenz gauge metric perturbation and its $r_0$ derivative: odd-sector} \label{Lorst} 
				
	Using the decomposition of $\h$ in \eqn{oddperturbation} and \eqn{evenperturbation} and following Berndtson's prescription \cite{Berndtson:2007gsc}, the homogeneous odd-sector components of $\hL$ for $\omega \neq 0$ and $l \geq 2$ are given by the radial fields:
\begin{align}
	h_{0}(r)&=\frac{1}{i \omega}\left(\psi_{1}+\frac{2 \lambda}{3} \psi_{2}\right), \label{einsteinodd1}\\
	h_{1}(r)&\label{einsteinodd2}=\frac{1}{(i \omega)^{2}}\bigg(-\frac{2 \lambda}{3} \psi_{2}^{\prime}+\frac{2}{r} \psi_{1}-\frac{2 \lambda}{3 r} \psi_{2}-\psi_1^{\prime}\bigg),\\
	h_{2}(r)&=\frac{1}{(i \omega)^{2}}\bigg(r f~ \psi_{2}^{\prime}+\psi_{1}+\frac{(3+2 \lambda) r-6M}{3 r} \psi_{2}\bigg), \label{einsteinodd3}
\end{align}
where $\psi_1$ and $\psi_2$ refer to the odd-sector solutions to \eqn{RWeqn2} with a RW potential, and whose sources $S_1$ and $S_2$ are given by \eqn{S1RW} and \eqn{S2RW}, respectively. Substituting the solutions to $\psi_s$ and their radial derivatives from \Cref{psiretsec} into Eqs.~\eqref{einsteinodd1}-\eqref{einsteinodd3}, we calculate the inhomogeneous Lorenz gauge metric perturbation for quasicircular orbits in the odd-sector.  

Note that away from the world-line, we can substitute the inhomogenous RW fields into Eqs.~\eqref{einsteinodd1}-\eqref{einsteinodd3} to obtain the inhomogeneous solution to the odd-sector metric perturbation. On the world-line however, additional source terms that appear in the inhomogeneous expressions for Eqs.~\eqref{einsteinodd1}-\eqref{einsteinodd3}, which can be found in Ref.~\cite{Berndtson:2007gsc}, all cancel with any distributional terms arising from radial derivatives of the RW fields. Therefore, inhomogeneous solutions to $h_0$, $h_1$ and $h_2$ in the odd-sector are mode-by-mode $C^0$ differentiable over the entire domain, as required in the Lorenz gauge.

We now wish to calculate $\hLr$ in the odd-sector, which we shall do so by making use of the gauge transformation in \eqn{gaugetransformeqn}, and taking its derivative with respect to $r_0$, or equivalently taking the $r_0$ derivative of Eqs.~\eqref{einsteinodd1}-\eqref{einsteinodd3}. 
As such, $\hLr$ in the odd-sector will consist of the RW master functions, their slow-time derivatives and both their radial derivatives. 
For example, taking the derivative of \eqn{einsteinodd1} with respect to $r_0$, we obtain: 
	\begin{equation}
		h_{0,r_0}(r)=\frac{1}{i \omega}\left(\dpsi_{1}+\frac{2 \lambda}{3} \dpsi_{2}\right) - \frac{h_0}{\omega}\omega_{,r_0},
	\end{equation}
	which can be calculated by substituting inhomogeneous solutions to $\psi_s$ and  $\dpsi_{s}$ from \Cref{psirretsec}, away from the world-line and matching at the particle.
	
	\subsection{Calculating the Lorenz gauge metric perturbation and its $r_0$ derivative: even-sector} \label{LorstEven} 

For $\omega \neq 0 $ and $l \geq 2$, the even-sector components of $\hL$, given by the radial fields $h_0$, $h_1$, $H_0$, $H_1$, $H_2$, $K$ and $G$ can be written in terms of the fields $\psi_{2\text{Z}}$, $\psi_1$, $\psi_0$, $\psi_{0b}$ and their radial derivatives, in addition to the gauge field $\mtaf$ and its radial derivative. The explicit expressions for $\hL$ in the even-sector are much longer than the odd-sector expressions, so we give them in \Cref{BEven}. The fields $h_0$, $h_1$ and $\psi_1$ should not be confused with those from the odd-sector. Here $\psi_1$, $\psi_0$, $\psi_{0b}$ refer to the even-sector solutions to \eqn{RWeqn} with a RW potential, whose sources are given by \eqn{Se1RW}-\eqref{Se0bRW} respectively. The field $\psi_{2\text{Z}}$ is also an even-sector solution to \eqn{RWeqn} with a Zerlli potential, whose source is given by \eqn{S2Z}. The additional gauge field $\mtaf$ is a contribution that comes from the trace free, $s = 0$ piece of the pure gauge term in \eqn{gaugetransformeqn} \cite{Berndtson:2007gsc, Dolan:2021ijg}. 
	The origin of $\mtaf$ is elucidated by the framework of Ref.~\cite{Dolan:2021ijg}. 
	A part of the Lorenz gauge metric perturbation that is pure gauge can be constructed by a spin-weight $s = 0 $ gauge vector in the following way:
	\begin{equation}
		h_{\alpha \beta}^{1\text{L} (s=0)}=-2 \xi_{(\alpha ; \beta)}^{(s=0)}.
	\end{equation}
	The part of this gauge vector that does not contain the trace of the metric perturbation is related to $\mtaf$. 
	As such, $\mtaf$ is a `pure gauge' field. 
	
	Similarly to the odd-sector, inhomogeneous solutions to $\hL$ are constructed in the even-sector using expressions in \Cref{BEven}, and are $C^0$ differentiable as required. Even-sector solutions to $\hLr$ are then also constructed similarly to those in the odd-sector.

	\subsection{Calculating $\mathbf{\mtaf}$}\label{mtafsec}
	
	The equation governing $\mtaf$ is given by \cite{Berndtson:2007gsc}:
	\begin{equation}
		\mathcal{L}_0 M_{2af}(r)= f(r)~\psi_0(r), \label{mtaf}
	\end{equation}
	where $\mathcal{L}_0$ is the operator from \eqn{RWeqn} with a RW potential and spin-weight $s=0$. 
	\Cref{mtaf} is exactly the RW equation, with a source that now contains $\psi_0$ and is thus unbounded.
	The form of \eqn{mtaf} means that we can again tackle it with the method of partial annihilators. 
	Applying an additional $\mathcal{L}_0$ operator to \eqn{mtaf} yields:
	\begin{equation}
		\mathcal{L}_0\left(\frac{1}{f} \mathcal{L}_0 \mtaf \right) = S_0, \label{mtaf2}
	\end{equation}
	where we have made use of the fact that $\mathcal{L}_0 \psi_0 = S_0$, with the source $S_0$ provided in \eqn{S0RWeven}.
	As an inhomogeneous, fourth-order ODE, \eqn{mtaf2} will have four independent homogeneous solutions.
	Two of these solutions will be $\psi_0^\pm$ from \Cref{psiretsec}.
	As the fourth-order Eq.~\eqref{mtaf2} does not have a known MST-type solution, the other two homogeneous solutions must be obtained by numerical integration starting with appropriate boundary conditions at finite radii.
	Series expansions of the homogeneous solutions near infinity and the horizon are given by:
	\begin{align}
		\mtaf^{h4,+}(r) &= r e^{i \omega r_*} \sum _{i=0}^{n_{\text{max}}}\frac{A^{h4,+}_i}{(r \omega)^i} \bigg\vert_{r = r_{\text{out}}},	\label{eq:m2afouth4+}			\\
		\mtaf^{h4,-}(r) &= f(r) e^{-i \omega r_*} \sum _{i=-1}^{n_{\text{max}}} f(r)^i A^{h4,-}_i\bigg\vert_{r = r_{\text{in}}}.			\label{eq:m2afouth4-}
	\end{align}
	The coefficients $A_i^{h4,\pm}$, not to be confused with those from \Cref{sec:RWZ_BCs}, are derived by substituting the above ans\"atze into \eqn{mtaf2} and matching the coefficients of $r$ and $f(r)$.
	For brevity we have suppressed some indices and functional dependence of the coefficients in the series expansion which depend on the parameters $l$, $m$ and $r_0$, similarly to those for $\phi_s$. 
	The coefficients $A_i^{h4,\pm}$ obey recursion relations provided in the supplementary material \cite{suppmat}, and are distinct to those defined in \eqn{eq:RWout} and \eqn{eq:RWin}. 
	For \eqn{eq:m2afouth4+} if we set $A_0^{h4,+} = 1$ and $A_1^{h4,+} = 0$ we find the series expansion satisfies \eqn{mtaf2} but not $\mathcal{L}_0 \mtaf^{h4,+}=0$ and thus we know we have found another linearly independent homogeneous solution.
	If we set $A_0^{h4,+} = 0$ and $A_1^{h4,+} = 1$ we recover the boundary conditions for $\psi_0^+$  given in \eqn{eq:psi+_numBC}.
	Similarly for \eqn{eq:m2afouth4-} if we set $A_{-1}^{h4,-} = 1 $ and $A_{0}^{h4,-} = 0 $ we find a solution that satisfies \eqn{mtaf2} but not $\mathcal{L}_0 \mtaf^{h4,-}=0$.
	If we set $A_{-1}^{h4,-} = 0 $ and $A_{0}^{h4,-} = 1 $ we recover the boundary conditions for $\psi_0^-$ given in \eqn{eq:psi-_numBC}. 
	
	Following the same procedure as in \Cref{RW} when deriving $\psi_s$ and $\dpsi_{s}$, we can write the inhomogeneous solution to \eqn{mtaf} as: 
	\begin{align}
			\mtaf(r) &= \left(c^{h2,-} \psi_0^-(r) + c^{h4,-} \mtaf^{h4,-}(r)\right)\Theta(r_0 - r) \nonumber  \\ 
		& +\left(c^{h2,+} \psi_0^+(r) + c^{h4,+} \mtaf^{h4,+}(r)\right)\Theta(r - r_0), \label{mtafret}
	\end{align}
	where the $c^{h2/4,\pm}$ are distinct from those in Eq.~\eqref{psirret}.
	Substituting \eqn{mtafret} into \eqn{mtaf2} and using the identity in \eqn{deltaeqn}, we match the coefficients of the Dirac delta functions and their radial derivatives with those in $S_0$ and solve the linear system of equation for $c^{h2/4,\pm}$.
	As before we can write this in the following form:
	\begin{align}
		\mathbf{C} = \mathbf{\Phi}^{-1}\cdot\mathbf{J},	\label{eq:CsFromJumpsM2af}
	\end{align}
	where $\mathbf{C} =(c^{h2,-},c^{h4,-},c^{h2,+},c^{h4,+})$ and:
	\begin{equation}
		\mathbf{\Phi}=\left(\begin{array}{cccc}
		{-\psi_0^{-}}				&{-\mtaf^{h4,-}}					&{\psi_0^{+}}				&{\mtaf^{h4,+}}\\
		{-\partial_{r}\psi_0^{-}}	&{-\partial_{r}\mtaf^{h4,-}}		&{\partial_{r}\psi_0^{+}}	&{\partial_{r}\mtaf^{h4,+}}\\
		{-\partial^2_{r}\psi_0^{-}}	&{-\partial^2_{r}\mtaf^{h4,-}}	&{\partial^2_{r}\psi_0^{+}}	&{\partial^2_{r}\mtaf^{h4,+}}\\
		{-\partial^3_{r}\psi_0^{-}}	&{-\partial^3_{r}\mtaf^{h4,-}}	&{\partial^3_{r}\psi_0^{+}}	&{\partial^3_{r}\mtaf^{h4,+}}
		\end{array}\right)_{r=r_0}.
	\end{equation}
	As the source $S_0$ contains only Dirac delta functions with no radial derivatives, the vector $\mathbf{J}$ has only one non-zero component such that $\mathbf{J} = (0,0,0, J_3)$, where:
	\begin{align}
		J_3 = \frac{-16\pi\sqrt{r_0(r_0-3M)}}{(r_0 - 2M)^2}Y^*_{lm}(\pi/2,0).
	\end{align}	
	
	\subsection{Calculating the $r_0$ derivative of $\mtaf$} \label{mtafrsec}
	
	To compute slow-time derivatives of the Lorenz gauge metric perturbation we must also compute $\mtafr$.
	By taking an  $r_0$ derivative of \eqn{mtaf} we obtain:
	\begin{equation}
		\mathcal{L}_0 \mtafr= f \dpsi_{0} - 2 \omega \omega_{,r_0} \mtaf, \label{noncompact2}
	\end{equation}
	which also has a unbounded source. 
	In fact both terms on the right hand side of \eqn{noncompact2} are unbounded as they are non-zero everywhere from the horizon to infinity. 
	We can choose to write $\mtafr$ as a linear combination of two fields such that:
	\begin{equation}
		\mtafr(r) = \chi_1(r) + \chi_2(r),		\label{eq:dr0M2afchi1chi2}
	\end{equation}
	with
	\begin{align}
		\mathcal{L}_0 \chi_1 & = f\dpsi_{0}, \label{chi1}\\
		\mathcal{L}_0 \chi_2 & = -2 \omega \omega_{,r_0}\mtaf. \label{chi2}
	\end{align}
	As both of these equations have unbounded sources, we turn once again to the method of partial annihilators. 
	By making use of \eqn{RWeqn2}, \eqn{noncompact1} and \eqn{mtaf}, we obtain two sixth-order ODEs with distributional sources:
	\begin{align}
		\mathcal{L}_0^2\bigg(\dfrac{1}{f} \mathcal{L}_0 \chi_1 \bigg) &= \mathcal{L}_0 S_{0,r_0} - 2 \omega \omega_{,r_0} S_0, \label{eq:chi16thorder}\\
		\mathcal{L}_0\bigg(\dfrac{1}{f} \mathcal{L}_0^2 \chi_2\bigg)  &= - 2 \omega \omega_{,r_0} S_0, 							\label{eq:chi26thorder} 
	\end{align}
	which we shall also solve via the method of variation of parameters. 
	Both \eqn{eq:chi16thorder} and \eqn{eq:chi26thorder} have six independent homogeneous solutions.
	For Eq.~\eqref{eq:chi16thorder}, four of these are given by $\psi_0^\pm$ and $\mtaf^{h4,\pm}$.
	The final pair we will denote by $\chi_1^{h6,\pm}$ and these satisfy the sixth-order equation but neither the second-order equation $\left(\mathcal{L}_0\chi_1^{h6,\pm}\neq 0\right)$ nor the fourth-order equation $\left(\mathcal{L}_0(1/f\mathcal{L}_0)\chi_1^{h6,\pm}\neq 0\right)$.
	For \eqn{eq:chi26thorder}, four of the homogeneous solutions are given by $\psi_0^\pm$ and $\dpsi_0^\pm$.
	Similarly we will denote the final pair by $\chi_2^{h6,\pm}$.

	For our numerical scheme we use the following series expansion of the $\chi_{k}^{h6,\pm}$ fields at finite radii as boundary conditions:
	\begin{align}
		\chi_{k}^{h6,+}(r_\text{out}) &= r^2 e^{i \omega r_{*}}\sum _{i=0}^{n_{\text{max}}} \frac{A^{h6,+}_{k,i}+B^{h6,+}_{k,i}\log(r)}{(r \omega )^{i}}\bigg \vert_{r = r_{\text{out}}},\label{eq:chiout}\\
		\chi_{k}^{h6,-}(r_\text{in}) &= e^{-i \omega r_*}\hspace{-0.2cm} \sum _{i=0}^{n_{\text{max}}} \hspace{-0.05cm}[A^{h6,-}_{k,i} \hspace{-0.1cm}+\hspace{-0.05cm}B^{h6,-}_{k,i}\log(f(r))]f(r)^i\big \vert_{r = r_{\text{in}}}, \label{eq:chiin}
	\end{align}
	where $k \in\{1,2\}$.
	The recursion relations for $A^{h6,\pm}_{k,i}$ and $ B^{h6,\pm}_{k,i}$ are derived by substituting the above ans\"atze into \eqn{eq:chi16thorder} and \eqn{eq:chi26thorder}.
	Similarly to those for $\mtaf$, the recursion relations depend on the parameters $l, m$ and $ r_0$.
	These recurrence relations are provided in the supplementary material \cite{suppmat}.
	
	For $\chi_1^{h6,+}$ the three undetermined coefficients are $A^{h6,+}_{1,1}, A^{h6,+}_{1,2}$ and $B^{h6,+}_{1,1}$.
	The field equations also enforce that $B^{h6,+}_{0,0} = 0$, $A^{h6,+}_{1,0} = B^{h6,+}_{1,1}/4$ and $B^{h6,+}_{1,2} = i(l(l+1)+1+4i\omega)B^{h6,+}_{1,1}/{2\omega}$.
	Setting $A^{h6,+}_{1,1} = B^{h6,+}_{1,1} = 0$ and $A^{h6,+}_{1,2} = 1$ we find the series in Eq.~\eqref{eq:chiout} approximates a solution that satisfies Eq.~\eqref{eq:chi16thorder} but not the ODEs with operators $\mathcal{L}_0$ or $\mathcal{L}_0(1/f\mathcal{L}_0)$.
	Thus $\chi_1^{h6,+}$ is a linearly independent from the other two bases $\{\psi_0^+, \mtaf^{h4,+}\}$. 
	For $\chi_1^{h6,-}$ the three undetermined coefficients are $A^{h6,-}_{1,0}, A^{h6,-}_{1,1}$ and $B^{h6,-}_{1,1}$.
	The field equation also sets $B^{h6,-}_{1,0} = 0$.
	Setting $A^{h6,-}_{1,0} = A^{h6,-}_{1,1} = 0$ and $B^{h6,-}_{1,1} = 1$ we find the series in Eq.~\eqref{eq:chiin} approximates a solution that satisfies Eq.~\eqref{eq:chi16thorder} but neither the second- nor fourth-order ODEs.
	This demonstrates that we have found another linearly independent homogeneous solution.
	
	For $\chi_2^{h6,+}$, the three undetermined coefficients are $A^{h6,+}_{2,0}, A^{h6,+}_{2,1}$ and $A^{h6,+}_{1,1}$.
	The field equations \eqn{eq:chi16thorder} and \eqn{eq:chi26thorder} also enforce that $B^{h6,-}_{2,0} = B^{h6,-}_{2,1} = 0$ and $B^{h6,-}_{2,1} = -i(4+l+l^2)A^{h6,+}_{2,0}/\omega + 2^{h6,+}_{2,1}$.
	Setting $A^{h6,+}_{2,0} = 1 $ and $A^{h6,+}_{2,1} = A^{h6,+}_{2,2} = 0$ we find the series in Eq.~\eqref{eq:chiout} approximates a solution that satisfies Eq.~\eqref{eq:chi26thorder} but not the ODEs with operators $\mathcal{L}_0$ or $\mathcal{L}^2_0$.
	Thus $\chi_2^{h6,+}$ is linearly independent from the other two bases $\{\psi_0^+, \dpsi^{h4,+}\}$. 
	For $\chi_2^{h6,-}$ the three undetermined coefficients are $A^{h6,-}_{2,0}, A^{h6,-}_{2,1}$ and $B^{h6,-}_{2,0}$.
	The field equation also enforces that $B^{h6,-}_{2,1} = i(1+l+l^2)B^{h6,-}_{2,0}/(i+4\omega)$.
	Setting $A^{h6,-}_{2,0} = A^{h6,-}_{2,1} = 0$ and $B^{h6,-}_{2,0} = 1$ we find the series in Eq.~\eqref{eq:chiin} approximates a solution that satisfies Eq.~\eqref{eq:chi26thorder} but not the second- or fourth-order ODEs. This demonstrates that we have found the final linearly independent homogeneous solution.

	We can then write the retarded solutions to \eqn{eq:chi16thorder} and \eqn{eq:chi26thorder} in the form:
	\begin{align}
		\chi_1(r) &= \chi_1^-(r)\Theta(r_0-r) + \chi_1^+(r)\Theta(r-r_0), \label{eq:chi1ret}\\
		\chi_2(r) &= \chi_2^-(r)\Theta(r_0-r) + \chi_2^+(r)\Theta(r-r_0), \label{eq:chi2ret}
	\end{align}
	where
	\begin{align}
		\chi_1^\pm(r) &= \left(c^{h2,\pm} \psi_0^\pm(r) + c^{h4,\pm} \mtaf^{h4,\pm}(r) + c^{h6,\pm}\chi_1^{h6,\pm}(r)\right),			 \\
		\chi_2^\pm(r) &= \left(c^{h2,\pm} \psi_0^\pm(r) + c^{h4,\pm} \dpsi_0^{h4,\pm}(r) + c^{h6,\pm}\chi_2^{h6,\pm}(r)\right).
	\end{align}
	The coefficients in the above equations can be found by substituting Eqs.~\eqref{eq:chi1ret} and \eqref{eq:chi2ret} into Eqs.~\eqref{eq:chi16thorder} and Eqs.~\eqref{eq:chi26thorder}, respectively, and matching the coefficients of the delta functions and their derivatives.
	For $\chi_1$ we can once again write the resulting system of equations in the now familiar form:
	\begin{align}
		\mathbf{C} = \mathbf{\Phi}^{-1}\cdot\mathbf{J},	\label{eq:CsFromJumps6thOrder}
	\end{align}
	where now $\mathbf{C} =(c^{h2,-},c^{h4,-},c^{h6,-},c^{h2,+},c^{h4,+},c^{h6,+})$ and:
	\begin{widetext}
	\begin{equation}
		\mathbf{\Phi}=\left(\begin{array}{cccccc}
		{-\psi_0^{-}}				&{-\mtaf^{h4,-}}				&-\chi_1^{h6,-}					&{\psi_0^{+}}				&{\mtaf^{h4,+}}		 			&\chi_1^{h6,+}				\\
		{-\partial_{r}\psi_0^{-}}	&{-\partial_{r}\mtaf^{h4,-}}	&-\partial^2_r\chi_1^{h6,-}		&{\partial_{r}\psi_0^{+}}	&{\partial_{r}\mtaf^{h4,+}}		&\partial_r\chi_1^{h6,+} 	\\	
		{-\partial^2_{r}\psi_0^{-}}	&{-\partial^2_{r}\mtaf^{h4,-}}	&-\partial^3_r\chi_1^{h6,-}		&{\partial^2_{r}\psi_0^{+}}	&{\partial^2_{r}\mtaf^{h4,+}}	&\partial^2_r\chi_1^{h6,+}	\\
		{-\partial^3_{r}\psi_0^{-}}	&{-\partial^3_{r}\mtaf^{h4,-}}	&-\partial^4_r\chi_1^{h6,-}		&{\partial^3_{r}\psi_0^{+}}	&{\partial^3_{r}\mtaf^{h4,+}}	&\partial^3_r\chi_1^{h6,+}	\\
		{-\partial^4_{r}\psi_0^{-}}	&{-\partial^4_{r}\mtaf^{h4,-}}	&-\partial^4_r\chi_1^{h6,-}		&{\partial^4_{r}\psi_0^{+}}	&{\partial^4_{r}\mtaf^{h4,+}}	&\partial^4_r\chi_1^{h6,+}	\\
		{-\partial^5_{r}\psi_0^{-}}	&{-\partial^5_{r}\mtaf^{h4,-}}	&-\partial^5_r\chi_1^{h6,-}		&{\partial^5_{r}\psi_0^{+}}	&{\partial^5_{r}\mtaf^{h4,+}}	&\partial^5_r\chi_1^{h6,+}	
		\end{array}\right)_{r=r_0}. \label{eq:Phi6thOrder}
	\end{equation}
	\end{widetext}
	For $\chi_1$, the four non-zero components of the vector $\mathbf{J}$ are given in Appendix \ref{apdx:Js_chi1}. For $\chi_2$, the matrix $\mathbf{\Phi}$ is the same in Eq.~\eqref{eq:Phi6thOrder} except $\mtaf^{h4,\pm}$ is replaced by $\dpsi_0^{h4,\pm}$ and $\chi_1^\pm$ is placed by $\chi_2^\pm$.
	As the source for $\chi_2$ contains only Dirac delta functions with no radial derivatives, the vector $\mathbf{J}$ has only one non-zero component and we find that $\mathbf{J} = (0, 0, 0, 0, 0, J_5)$ where:
	\begin{align}
		J_5 = -\frac{48 \pi m^2 M \sqrt{r_0-3 M}}{r_0^{3/2} (r_0-2 M)^4}Y^*_{lm}(\pi/2,0).
	\end{align}

	\subsection{Transforming to the Barack-Lousto-Sago basis}\label{sec:BSL}
		
	We now have everything we need to compute $\hL$ and $\hLr$ away from the world-line for quasicircular orbits.
	The setup described above computes the perturbation in Berndtson's spherical harmonic basis given by Eqs.~\eqref{eq:Wlm}-\eqref{eq:Ynorm}.
	For comparison with prior work \cite{Barack:2005nr,Akcay:2010dx} and for input into current second-order calculations \cite{Pound:2019lzj,Miller:2020bft,Warburton:2021kwk,Wardell:2021fyy} we will transform to the BLS basis which is used in Eq.~\eqref{eq:LorenzFourierDecomp} and given explicitly in Refs.~\cite{Barack:2007tm, Barack:2005nr}.
	By comparing the components of $\hL$ in each of \cite{Berndtson:2007gsc} and \cite{Barack:2007tm} we can find the radial Lorenz gauge metric perturbation components in the BLS basis in terms of Berndtson's variables.
	For the even-sector these are: 
	\begin{subequations}
	\begin{align}
		\bar{h}^{(1)}_{lm} & = r f (H_0^{lm} + H_2^{lm}),\\
		\bar{h}^{(2)}_{lm} & = 2 r f H_1^{lm},\\
		\bar{h}^{(3)}_{lm} & = 2 r K^{lm},\\
		\bar{h}^{(4)}_{lm} & = 2 l(l+1) h_0^{lm},\\
		\bar{h}^{(5)}_{lm} & = 2 l(l+1) f h_1^{lm},\\
		\bar{h}^{(6)}_{lm} & =  r (H_0^{lm}-H_2^{lm}),\\
		\bar{h}^{(7)}_{lm} & = 2 r (l-1)l(l+1)(l+2) G^{lm},
	\end{align}
	\text{and for the odd-sector:}
	\begin{align}
		\bar{h}^{(8)}_{lm} & = 2 l(l+1) h_0^{lm},\\
		\bar{h}^{(9)}_{lm} & = 2 l(l+1) f h_1^{lm},\\
		\bar{h}^{(10)}_{lm}& = -\dfrac{2(l-1)l(l+1)(l+2)}{r}h_2^{lm}.
	\end{align}
	\end{subequations}
		
	\subsection{Numerical Results}\label{sec:Lorenz_results}

\begin{figure}%[h]
		\includegraphics[width=0.48\textwidth]{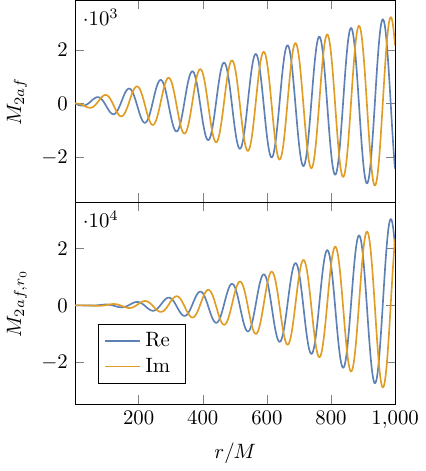}
		\caption{Real and imaginary parts of $\mtaf$ (top panel) and $\mtafr$ (bottom panel) for the $(l,m)=(2,2)$ mode with $r_0 = 10M$. 
		The amplitudes of the fields grow with $r$ and $r^2$ for large $r$ respectively, as determined by the asymptotic boundary conditions, \eqn{eq:m2afouth4+} and \eqn{eq:chiout}.
		Similar results are obtained for other even multiple modes and different values of $r_0$.}\label{fig:M2afreim}
	\end{figure}

\begin{figure}%[!h]
		\includegraphics[width=0.48\textwidth]{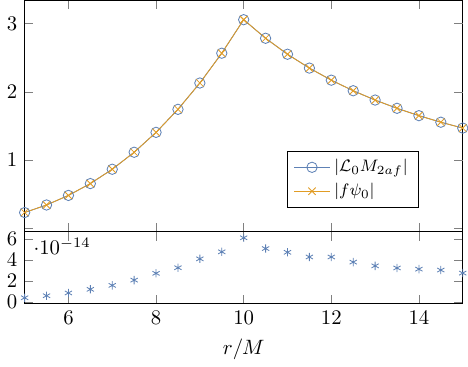}
		\caption{
			The top panel shows the absolute values of the left- and right-hand sides of \eqn{mtaf}, for $\mtaf$ given the solution computed using the fourth-order equation, \eqn{mtaf2}.
			The data shown is for $r_0 = 10 M$ and $(l,m) = (2,1)$.
			The lower panel shows that the absolute difference between the data sets for the left- and right-hand side of \eqn{mtaf} is near machine precision.
			}\label{fig:M2afCheck}
	\end{figure}

\begin{figure}%[h]
		\includegraphics[width=0.48\textwidth]{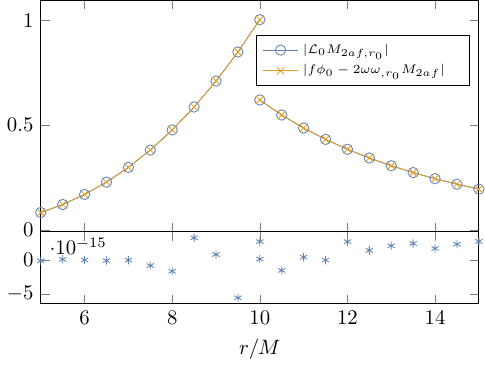}
			\caption{
			The top panel shows the absolute values of the left- and right-hand sides of \eqn{noncompact2}, for $\mtafr$ given the solution computed using Eq.~\eqref{eq:dr0M2afchi1chi2} and the two sixth-order equations, \eqn{eq:chi16thorder} and \eqn{eq:chi26thorder}.
			The data shown is for $r_0 = 10 M$ and $(l,m) = (2,2)$ . 
			The lower panel shows the absolute difference between the data sets for the left- and right-hand side of \eqn{noncompact2} is near machine precision.
			}\label{fig:dr0M2afCheck}
	\end{figure}

	In this section we present our numerical results for $\hL$ and $\hLr$ and a variety of checks that provide confidence in our results. We begin by showing results for $\mtaf$ and $\mtafr$ in Fig.~\ref{fig:M2afreim}.
	As a check on the implementation of our partial annihilator scheme for $\mtaf$ we verify that our solution to the fourth-order equation with a distributional source, \eqn{mtaf2}, satisfies the original second-order equation \eqn{mtaf} with an unbounded source.
	Similarly, we check that our solution for $\mtafr$, computed as the sum of the solutions of two sixth-order equations, \eqn{eq:chi26thorder} and \eqn{eq:chi16thorder}, satisfies the original second-order equation, \eqn{noncompact2}, with an unbounded source.
	Figures \ref{fig:M2afCheck} and \ref{fig:dr0M2afCheck} show that the relevant second-order equations are satisfied to near machine precision.
	
\begin{figure}%[H]
	\includegraphics[width=0.48\textwidth]{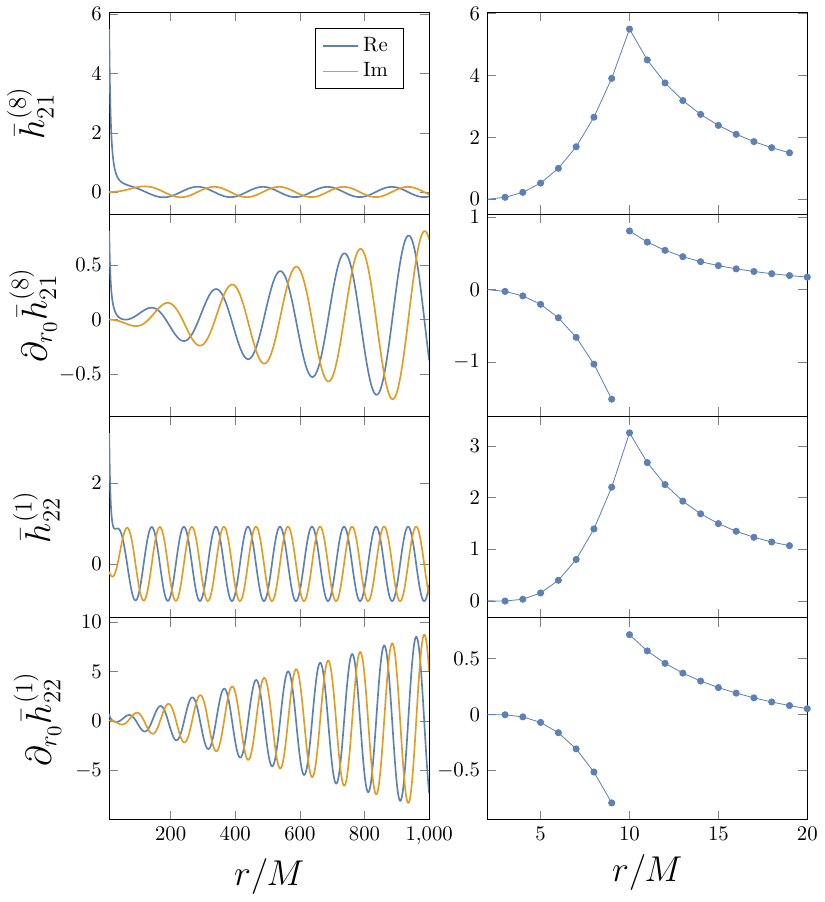}
	\caption{Left panels: real and imaginary parts of the metric perturbation components in the BSL basis, $\bar{h}^{(8)}_{21}$ (top) and $\partial_{r_0}\bar{h}^{(8)}_{21}$ (second), $\bar{h}^{(1)}_{22}$ (third) and $\partial_{r_0}\bar{h}^{(1)}_{22}$ (bottom). 
	Plots of the odd-sector fields are for the $(l,m) = (2,1)$ mode and plots of the event-sector fields are for the $(l,m)=(2,2)$ mode. 
	In both sectors, a value of $r_0 = 10 M$ is chosen. 
	Results are similar for the other remaining Lorenz gauge fields and for different choices of $r_0$ and $(l,m)$. 
	Right panels: the same results as for the left panels are shown, except only the real component of the fields are given near the particle. We see $C^0$ differentiability behaviour for $\bar{h}^{(1)}_{22}$  and $\bar{h}^{(8)}_{21}$  as required.} \label{COMBINED}
\end{figure}

	With confidence in our solution for $\mtaf$ we can compute $\hL$ and find agreement with previously published results \cite{Akcay:2013wfa} to the accuracy of those results.
	We present sample results for both $\hL$ and $\hLr$ in Fig.~\ref{COMBINED}.

\begin{figure}[!h]
	\includegraphics[width=0.48\textwidth]{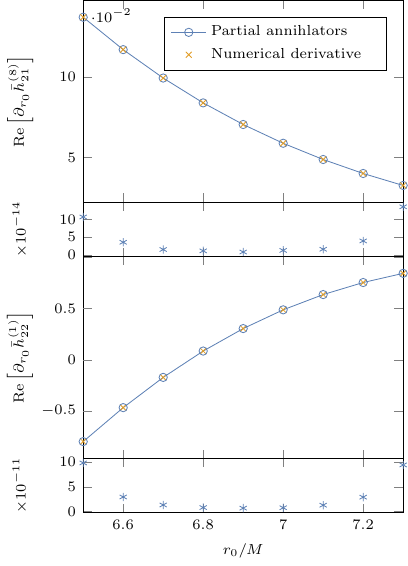}
	\caption{
	Comparison of numerical results for $\hLr$ computed using our partial annihilator method and by taking a numerical derivative. 
	The top panel shows the real part of $\partial_{r_0}\bar{h}^{(8)}_{21}$ and the third panel shows the real part of $\partial_{r_0}\bar{h}^{(1)}_{22}$ at a fixed field point $r=50$ as a function of $r_0$.
	The second and forth panels show that the absolute difference between numerically computing the $r_0$ derivative and computing the $r_0$ derivative using our partial annihilators method is near machine precision.
	We find similar results for all other remaining Lorenz gauge fields, their radial derivatives as well as for a different choice of modes, fixed field point and set of radii $r_0$.}\label{fig:hLrComparison}
\end{figure}

As a check on our results for $\partial_{r_0}\bar{h}^{(i)}_{lm}$ we compare them with a numerically computed $r_0$ derivative of $\bar{h}^{(i)}_{lm}$. In our check we computed data from our partial annihilator method at $r = 50 M$ with $r_0$ ranging from $6.5 M$ to $7.3M$ in steps of $0.1M$. The numerical derivatives are obtained by interpolating data for the metric perturbation for different values of $r_0$, keeping the field point $r$ constant. For the numerical derivatives data was computed at $r=50M$ for $r_0$ ranging from $6.5M$ to $7.5M$ in steps of $0.1M$. The wider $r_0$ range was used for the interpolant data to avoid error at the end points of the interpolation. We present the result of the comparison in Fig.~\ref{fig:hLrComparison} where we find near machine precision agreement between our results for $\partial_{r_0}\bar{h}^{(8)}_{21}$ and $\partial_{r_0}\bar{h}^{(1)}_{22}$ and the numerical $r_0$ derivatives of $\bar{h}^{(8)}_{21}$ and $\bar{h}^{(1)}_{22}$ respectively. 

It is instructive again to consider the efficiency of the partial annihilator approach verses taking the numerical derivative. One significant new calculation introduced in the gauge transformation in the even sector is the calculation of $M_{2af,r_0}$. Using the partial annihilator method we must solve the left-hand side of Eq.~\eqref{eq:chi16thorder} 6 times and Eq.~\eqref{eq:chi26thorder} 4 times (two less as two the bases are the same as for Eq.~\eqref{eq:chi16thorder}), giving a total of 10 ODE evaluations. Computing $\mtaf$ requires solving the left-hand side of Eq.~\eqref{mtaf2} 4 times and to numerically calculation its $r_0$ derivative to a relative precision of $\sim10^{-15}$ at, e.g., $r_0=7.5M$ we find we must evaluate $\mtaf$ 8 times at equally spaced radii between $r_0 = 7.5 \pm0.1M$ and fit to an eighth-order polynomial. This gives a total of $4\times 8 = 32$ ODE evaluations. This means our approach is $32/10=3.2$ times faster than numerically computing the derivative of $M_{2af,r_0}$. In practice, when computing $\hLr$ the speed up is about twice this as four of the homogeneous solutions for Eqs.~\eqref{eq:chi16thorder} and \eqref{eq:chi26thorder}, namely $\psi_0^\pm$ and $\phi_0^\pm$, will already be computed in other parts of the calculation. 

For second-order calculations, where $\hLr$ is needed at thousands of field points using numerical derivatives is even less favourable as data for each field point $r$ needs to be calculated for a tens of values of $r_0$, fitted to a high-order polynomial and the linear coefficient extracted. As such, our method of calculating $\hLr$ is strongly favourable in this case.

	\section{Conclusions} \label{conclusion}
		
	In this paper we present a novel calculation for the $r_0$ derivative of the radiative modes of the first-order metric perturbation in the Lorenz gauge, on a Schwarzschild background, for circular orbits in the frequency domain. Our work can also be used to calculate the metric perturbation in the RW gauge automatically.
	Using the gauge transformation first derived in Ref.~\cite{Berndtson:2007gsc}, we obtain solutions to $\hL$ in terms of RWZ master functions and the gauge field $\mtaf$, in addition to their radial derivatives. 
	We also obtain $\hLr$ in terms of RWZ master functions, $\mtaf$, their derivatives with respect to $r_0$, in addition the radial derivatives of the listed fields. 
	The field equations for the $r_0$ derivative of the RWZ master functions and $\mtaf$ have sources which with unbounded support which is challenging for the standard variation of parameters method to tackle.
	To overcome this we use the method of partial annihilators \cite{Hopper:2012ty} which gives us higher-order differential equations but with distributional sources.
	While this work was being prepared for publication, a similar procedure for computing the RWZ master functions was sketched in Appendix C of Ref.~\cite{Mathews:2021rod}.
	In Secs.~\ref{sec:RW_results} and \ref{sec:Lorenz_results} we present the numerical results of our calculation and show they agree with results obtained by taking a numerical derivative of the relevant field.
	The results of this work are already being used to compute slow-time derivatives of the metric perturbation that feed into the source for second-order in the mass ratio calculations \cite{Miller:2020bft, Warburton:2021kwk, Wardell:2021fyy}.
	These form part of the recent calculations of the the second-order energy flux at infinity \cite{Warburton:2021kwk} and waveforms through second-order \cite{Wardell:2021fyy}.
	
	There are a variety of natural extensions to this work.
	Firstly, with the numerical results we already have, we could immediately construct the first-order metric perturbation and its $r_0$ derivative in the RW gauge.
	Secondly, the partial annihilator approach can also be applied to the Teukolsky formalism \cite{Teukolsky:1973ha}.
	This would allow for the slow-time derivative of the metric perturbation of a particle moving on either circular or spherical orbits around a Kerr black hole to be calculated in the radiation gauge \cite{Chrzanowski:1975wv,Kegeles:1979an,Wald:1978vm} following recent work by Green et al. \cite{Green:2019nam} and Toomani et al. \cite{Toomani:2021jlo}, who present regularised schemes on how to implement the second-order Teukolsky equation in Kerr. Alternatively, the same calculation could be done in the Lorenz gauge following the work of Dolan et al. \cite{Dolan:2021ijg}. 
	Unfortunately, it seems unlikely that the method of partial annihilators will be directly applicable to second-order perturbations to give a compact source.  
	The second-order perturbation to the Einstein tensor that appears in the second-order source is formed from quadratic combinations of the modes of the first-order metric perturbation and its derivatives and there is no obvious operator that would (partially) annihilate any pair of first-order modes, let alone all of them.
	
	Calculations for eccentric orbits are more challenging than circular orbits.
	The same partial annihilator method can be applied to eccentric orbit equations, arriving at sources which only have support within the radial libration region \cite{Hopper:2012ty}.
	For highly eccentric orbits it is likely that a time-domain approach would be preferable.
	In this approach, the derivative with respect to eccentricity or, e.g., the semi-major axis would only act on the source, leaving the time-domain operator unchanged.
	This suggests that already existing time-domain codes, e.g., \cite{Sundararajan:2007jg, Harms:2014dqa, Heffernan:2017cad}, could be quickly modified to calculate the slow-time derivatives for these orbits.
	
	We also note that whilst this work was being prepared, a new numerical approach to frequency-domain calculations of perturbations of black hole spacetimes was developed \cite{PanossoMacedo:2022fdi}.
	That work showed, using a scalar-field toy model, that using hyperboloidal compactified coordinates with a pseudo-spectral numerical scheme allows the $r_0$ derivative of the scalar field to be calculated efficiently.
	It would be interesting future work to compare the pros and cons of the partial annihilator method used in this work with the hyperboloidal approach.
	
	Finally, we note that up to $\mtaf$, $\hL$ can be written in terms of semi-analytic Mano, Suzuki and Takasugi (MST) expansions \cite{Mano:1996vt}. 
	The gauge field $\mtaf$ currently has no known MST-type expansion however, excluding analytic high-order PN calculations of $\hL$ similar to those in, e.g., \cite{Fujita:2012cm,Kavanagh:2015lva,Bini:2014ica,Munna:2022xts}.
	Research is currently being done to find a PN solution for $\mtaf$ \cite{Kavanagh:2021}, which would allow one to obtain a PN series solutions of the complete Lorenz gauge metric perturbation. \newline
		
	\begin{acknowledgments}
	We thank Adrian Ottewill and Barry Wardell for many helpful discussions.
	We thank Josh Matthews for his help in comparing numerical solutions to the RWZ master functions and in deriving the boundary conditions for the fourth-order Zerilli equation. 
	We thank Sam Dolan and Barry Wardell for their input in understanding $\mtaf$. We would also like to thank the anonymous reviewer of this paper for their thorough reading and many helpful comments.
	NW acknowledges support from a Royal Society - Science Foundation Ireland University Research Fellowship via grants UF160093 and RGF\textbackslash R1\textbackslash180022.
	This work also makes use of the Black Hole Perturbation Toolkit \cite{BHPToolkit}.
	\end{acknowledgments}
	
	\newpage
	\onecolumngrid
	\appendix
	\section{The metric perturbation in a tensor spherical harmonic basis} \label{tensorbasis}
	The source to $h^1_{\mu \nu}$, given by the stress-energy tensor $T^1_{\mu \nu}$ can be decomposed into a tensor spherical harmonic basis in the following way \cite{Berndtson:2007gsc}:
	\begin{equation}
	T^1_{\mu \nu}(t, r, \theta, \phi)=\sum_{l=0}^{\infty} \sum_{m=-l}^{l} \int_{-\infty}^{\infty} e^{-i \omega t}\left(T_{\mu \nu}^{o, l m}(\omega, r, \theta, \phi)+T_{\mu \nu}^{e, l m}(\omega, r, \theta, \phi)\right) d \omega.
	\end{equation}
	Similarly to $h^1_{\mu \nu}$ in \Cref{RW}, dependence on $\theta$, $\phi$ and $\omega$ is dropped. The odd- and even-sector stress-energy tensors can then be written as: 
	\begin{equation}
	T_{\mu \nu}^{o, l m}(r)=\left(\begin{array}{cccc}
	0 & 0 & S o_{02}^{l m}( r) \csc \theta \frac{\partial Y_{l m}}{\partial \phi} & -S o_{02}^{l m}( r) \sin \theta \frac{\partial Y_{l m}}{\partial \theta} \\
	* & 0 & S o_{12}^{l m}(r) \csc \theta \frac{\partial Y_{l m}}{\partial \phi} & -S o_{12}^{l m}( r) \sin \theta \frac{\partial Y_{l m}}{\partial \theta} \\
	* & * & -S o_{22}^{l m}( r) X_{l m} & S o_{22}^{l m}( r) \sin \theta W_{l m} \\
	* & * & * & So_{22}^{l m}(r) \sin ^{2} \theta X_{l m}
	\end{array}\right),\label{Todd}
	\end{equation}
	and
	\begin{equation}
	\begin{aligned}
	&T_{\mu \nu}^{e, l m}(r)= \\
	&\left(\begin{array}{cccc}
	S e_{00}^{l m}( r) Y_{l m} & S e_{01}^{l m}(r) Y_{l m} & S e_{02}^{l m}( r) \frac{\partial Y_{l m}}{\partial \theta} & S e_{02}^{l m}(r) \frac{\partial Y_{l m}}{\partial \phi} \\
	* & S e_{11}^{l m}(r) Y_{l m} & S e_{12}^{l m}(r) \frac{\partial Y_{l m}}{\partial \theta} & S e_{12}^{l m}(r) \frac{\partial Y_{l m}}{\partial \phi} \\
	* & * & U e_{22}^{l m}(r) Y_{l m} +S e_{22}^{l m}(r) W_{l m} & S e_{22}^{l m}( r) \sin \theta X_{l m} \\
	* & * & * & \sin ^{2} \theta\left(U e_{22}^{l m}(r) Y_{l m}\right. \left.-S e_{22}^{l m}( r) W_{l m}\right)
	\end{array}\right).
	\end{aligned}\label{Teven}
	\end{equation}
	
	\newpage
	\section{Sources for Regge-Wheeler and Zerilli master functions} \label{Sources}
	Here we provide expressions for the sources to the RWZ master functions from \eqn{RWeqn2} in terms of the components of $T_{\mu \nu}^1$. All of the expressions in this section are provided by reference \cite{Berndtson:2007gsc}. The components of $T_{\mu \nu}^1$ are given explicitly in \Cref{source components}. The following expressions are the sources for the odd-sector RW  master functions \cite{Berndtson:2007gsc}:
	\begin{align}
	S^{lm}_{1}(r)\label{S1RW} =~ & \frac{32 \pi  (\lambda +3) (r-2 M)^2 }{3 r^3}So^{l m}_{12}(r)+\frac{32 \pi  \lambda  (r-2 M) }{3 r^3}So^{l m}_{22}(r)+\frac{16 \pi  (r-2 M)^3 }{r^3}\partial_r So^{l m}_{12}(r)\\&+\frac{32 \pi  \lambda  (r-2 M)^2
	}{3 r^3}\partial_r So^{l m}_{22}(r),\nonumber\\\nonumber&\\
	S^{lm}_{2}(r) \label{S2RW} =~ &-\frac{16 \pi  f(r)^2}{r} So^{l m}_{12}(r)+\frac{32 \pi  \left(6 M^2-5 M r+r^2\right) }{r^4}So^{l m}_{22}(r)-\frac{16 \pi  f(r)^2}{r}\partial_rSo^{l m}_{22}(r).
	\end{align}
	The following expressions are the sources for the even-sector RWZ master functions. Here $S_2^{lm}$ refers to the source for the Zerilli master function. The remaining expressions are sources for the even-sector RW master function\cite{Berndtson:2007gsc}:
	\begin{align}
	S^{lm}_{2}(r) \label{S2Z}=~
	&\frac{16  \pi  M (2 M-r) (3 M-(\lambda +3) r)}{i \omega r (3 M+\lambda  r)^2}Se^{l m}_{01}(r)
	+\frac{8 \pi  (r-2 M)^2}{3 M+\lambda  r}Se^{l m}_{11}(r)\\&
	+\frac{16 \pi  (2 M-r) \left(6 M^2+(\lambda -3) M r+\lambda  (\lambda +1) r^2\right)}{i \omega r^2   (3 M+\lambda  r)^2} Se^{l m}_{02}(r)
	+\frac{16 \pi  (r-2 M)^2 }{r (3 M+\lambda  r)}Se^{l m}_{12}(r)	\nonumber\\&
	+\frac{32 \pi  (2 M-r)}{r^2} Se^{l m}_{22}(r)
	+\frac{8 \pi  (r-2 M)^2 }{i \omega  (3 M+\lambda  r)}\partial_r Se^{l m}_{01}(r)
	+\frac{16 \pi  (r-2 M)^2 }{i \omega r (3 M+\lambda  r)}\partial_r Se^{l m}_{02}(r),
	\nonumber\\\nonumber&\\
	S^{lm}_{1}( r) \label{Se1RW}=~ &\frac{16  \pi  r}{3 i \omega }Se^{l m}_{00}(r)+\frac{32 \pi  (r-2 M)^2}{3 (i\omega)^2 r^2} Se^{l m}_{01}(r)+\frac{64 \pi  M (2 M-r)
	}{3 (i\omega)^2 r^3}Se^{l m}_{02}(r)\\&-\frac{16 \pi  (r-2 M)^2}{3 i \omega r }Se^{l m}_{11}(r)+\frac{16 \pi  (r-2 M)^2}{3 i \omega r^2}Se^{l m}_{12}(r)+\frac{32 \pi  \lambda  (2 M-r)}{3
		i \omega  r^2 }Se^{l m}_{22}(r)\nonumber\\&+\frac{16 \pi  (2 M-r)}{i \omega r^2 }Ue^{l m}_{22}(r)-\frac{32 \pi  (r-2 M)^2 }{3 (i \omega)^2 r^2 }\partial_r Se^{l m}_{02}(r),\nonumber\\&\nonumber\\
	S^{lm}_{0}(r) \label{S0RWeven} =~ &-16 \pi  r Se^{l m}_{00}(r)+\frac{16 \pi  (r-2 M)^2}{r}Se^{l m}_{11}(r)+\frac{32 \pi  (r-2 M)}{r^2}Ue^{l m}_{22}(r),\\\nonumber&\\
	S^{lm}_{0b}(r) \label{Se0bRW} =~ &\frac{4 \pi  r \left((-6 \lambda -5) M+r \left(3 \lambda+2  (i\omega)^2r^2+2\right)\right)}{(i\omega)^2 (2 M-r)}Se^{l m}_{00}(r)\\&-\frac{8 \pi  \left(M \left(-2 \lambda +8
		(i\omega)^2r^2-3\right)-2  (i\omega)^2r^3+\lambda  r+r\right)}{(i\omega)^3 r}Se^{l m}_{01}(r)-\frac{16 \pi  (\lambda +1) \left(M-2 (i\omega)^2 r^3\right)
	}{(i\omega)^3r^2 }Se^{l m}_{02}(r)\nonumber\\&+\frac{4 \pi  (2 M-r) \left((14 \lambda +17) M-r \left(7 \lambda +2 (i\omega)^2 r^2+8\right)\right)}{(i \omega)^2 r}Se^{l m}_{11}(r)-\frac{8 \pi  (\lambda +1) (2 M-r)}{(i\omega)^2 r}Se^{l m}_{12}(r)\nonumber\\&+\frac{16 \pi  \lambda  (\lambda +1) (2 M-r)}{(i\omega)^2 r^2}Se^{l m}_{22}(r)-\frac{8 \pi  (6 \lambda +7) (2 M-r) }{(i\omega)^2 r^2}Ue^{l m}_{22}(r)\nonumber\\&-\frac{8 \pi  (\lambda +1) (2 M-r)}{(i\omega)^3 r}\partial_r Se^{l m}_{02}(r).\nonumber
	\end{align}	
	
	\newpage
	\section{Components of $T^1_{\mu\nu}$ in the frequency domain for quasicircular orbits} \label{source components}
	Here we provide leading-order, frequency domain expressions for the components of the stress-energy tensor from \eqn{Todd} and \eqn{Teven}, in the case of quasicircular, equatorial orbits. All sources are compactly supported on the particle's world-line and are provided by reference \cite{Berndtson:2007gsc}:
	
	\begin{align}
	So^{l m}_{02}(r)&=\mu u^t\frac{ f(r) \Omega_0}{l (l+1)}\delta (r-r_0)  \partial_{\theta}Y^*_{lm}\Big(\frac{\pi}{2},0\Big), \label{So02circ}\\
	So^{l m}_{12}(r)&=0,\\
	So^{l m}_{22}(r)&=-2 i m \mu u^t\frac{r^2 \Omega_0^2 }{2l (l+1)(l-1)(l+2)} \delta (r-r_0)  \partial_{\theta}Y^*_{lm}\Big(\frac{\pi}{2},0\Big),\\
	Se^{l m}_{00}(r)&=\mu u^t\frac{ f(r)^2}{r^2} \delta (r-r_0) Y^*_{lm}\Big(\frac{\pi}{2},0\Big),\\
	Se^{l m}_{01}(r)&=0, \\
	Se^{l m}_{11}(r)&=0,\\
	Ue^{l m}_{22}(r)&=\frac{1}{2} \mu u^t r^2 \Omega_0^2 \delta (r-r_0)  Y^*_{lm}\Big(\frac{\pi}{2},0\Big), \\
	Se^{l m}_{02}(r)&=i m\mu u^t \frac{ f(r) \Omega_0 }{l (l+1)}\delta (r-r_0)  Y^*_{lm}\Big(\frac{\pi}{2},0\Big),\\
	Se^{l m}_{12}(r)&=0, \\
	Se^{l m}_{22}(r)&=\mu u^t\frac{ r^2 \Omega_0^2(l(l+1)-2 \sc{m}^2)}{2l (\l+1)(l-1)(l+2)}\delta (r-r_0)Y^*_{lm}\Big(\frac{\pi}{2},0\Big), \label{Se22circ}
	\end{align}
	where the $t$ component of the four-velocity of the secondary is given by \cite{Berndtson:2007gsc}:
	\begin{equation}
	u^t = \sqrt{\frac{r_0}{r_0-3M}}.
	\end{equation}
	
	% \section{Properties of delta functions and their derivatives} \label{delta}
	% The following properties can be derived by taking repeated derivatives of the first rule, rearranging and applying previous rules:
	% \begin{align}
	% f(x) \delta(x - a) =& f(a) \delta(x - a)\\
	% f(x)\delta^{\prime}(x-a) =& -f^{\prime}(a) \delta(x-a)+f(a)\delta^{\prime}(x-a)\\
	% f(x)\delta^{\prime \prime}(x-a) =& f^{\prime \prime}(a) \delta(x-a)-2f^{\prime}(a)\delta^{\prime}(x-a)+f(a)\delta^{\prime \prime}(x-a)\\
	% f(x)\delta^{\prime \prime \prime}(x-a) =& -f^{\prime \prime \prime}(a) \delta(x-a)-3f^{\prime}(a)\delta^{\prime \prime}(x-a)+3f^{\prime \prime}(a)\delta^{\prime}(x-a)+f(a)\delta^{\prime \prime \prime}(x-a)\\
	% f(x)\delta^{(4)}(x-a) =& f^{(4)}(a) \delta(x-a)-4f^{\prime \prime \prime}(a)\delta^{\prime}(x-a)+6f^{\prime \prime}(a)\delta^{\prime \prime}(x-a)-4f^{\prime}(a)\delta^{\prime \prime \prime}(x-a)+f(a)\delta^{(4)}(x-a)\\
	% f(x)\delta^{(5)}(x-a) =& -f^{(5)}(a) \delta(x-a)+5f^{(4)}(a)\delta^{\prime}(x-a)-10f^{\prime \prime \prime}(a)\delta^{\prime \prime}(x-a)+10f^{\prime \prime}(a)\delta^{\prime \prime \prime}(x-a)\\&\nonumber-5f^{\prime}(a)\delta^{(4)}(x-a)+f(a)\delta^{(5)}(x-a)
	% \end{align}
	% where prime denotes derivatives with respect to $x$ and $a$ is a constant.
	
	\newpage
	\section{Coefficients of Dirac delta functions and the radial derivatives of Dirac delta functions for $S_2$ in the odd-sector} \label{b's}
	
	\begin{align}
	b_0(r_0) = & ~\frac{8 i m M \pi}{(l+2)(l+1)l(l-1)r_0^{10}}\left(\frac{r_0}{r_0-3M}\right)^{3/2}\big\{1080(l^2+l+1)M^4-6(246 l(l+1)-m^2+194)M^3r_0\\&\nonumber\\&\nonumber+(722l(l+1)-m^2+412)M^2r_0^2-3(49l(l+1)+16)M r_0^3+10l(l+1)r_0^4\big\} \partial_\theta Y_{lm}^*(\pi/2, 0),\\\nonumber &\\
	b_1(r_0) = & ~\frac{8 i m M \pi(2M-r_0) }{(l+2)(l+1)l(l-1)r_0^{10}}\left(\frac{r_0}{r_0-3M}\right)^{3/2}\big\{3240M^4+6(38l(l+1)+m^2-430)M^3 r_0\\&\nonumber\\&\nonumber+(-248l(l+1)+11m^2+604)M^2r_0^2+(87l(l+1)-4m^2-32)Mr_0^3-10l(l+1)r_0^4\big\}\partial_\theta Y_{lm}^*(\pi/2, 0),\\\nonumber &\\
	b_2(r_0) = &  ~\frac{8 i m M \pi(r_0-2M)^2}{(l+2)(l+1)l(l-1)r_0^{9}}\left(\frac{r_0}{r_0-3M}\right)^{3/2}\big\{996M^3+2(6l(l+1)+3m^2-260)M^2 r_0\\&\nonumber\\&\nonumber-(10l(l+1)+2m^2-49)Mr_0^2+2(l(l+1)+3)r_0^3\big\}\partial_\theta Y_{lm}^*(\pi/2, 0),\\\nonumber &\\\nonumber
	b_3(r_0) = & ~\frac{8 i m M \pi(2M - r_0)^3 }{(l+2)(l+1)l(l-1)r_0^{8}}\left(\frac{r_0}{r_0-3M}\right)^{3/2}\left(126M^2-47Mr_0+2r_0^2\right) \partial_\theta Y_{lm}^*(\pi/2, 0),\\\nonumber &\\
	b_4(r_0) = & -\frac{16 i m M \pi (r_0-2M)^4}{(l+2)(l+1)l(l-1)r_0^6}\sqrt{\frac{r_0}{r_0-3M}}\partial_\theta Y_{lm}^*(\pi/2, 0).
	\end{align}

	% \section{Coefficients of delta-functions and the radial derivatives of delta-functions for \eqn{LHS1}}\label{c's}
	% \eqn{LHS1}
	% \begin{align}
	% c_1(r_0) = &~\\
	% c_2(r_0) = &~\\
	% c_3(r_0) = &~\\
	% c_4(r_0) = &~\left\{\sum_{i=h2}^{h4} \left(c^{i,+}\psi^{i,+}_{s,r_0}- c^{i,-}\psi^{i,+}_{s,r_0}\right)f^4-10c^{\delta}\psi^{\delta}_{s,r_0}f^3f'\right\}\bigg\vert_{r=r_0}\\
	% c_5(r_0) = &~ c^{\delta}\psi^{\delta}_{s,r_0}(r_0)f(r_0)^4
	% \end{align}

	\newpage
	\section{Jump conditions} \label{apdx:Js}
	
	In this appendix we present the jump conditions required to construct the various fields that go into calculating $\hLr$. 
	The jump conditions for the $r_0$ derivative of the Zerilli master function are too long to present here so we including them in the supplemental material \cite{suppmat}.
	
	\subsection{Jump conditions in $\dpsi_2$ in the odd-sector}
	
	For the odd-sector the jump conditions in Eq.~\eqref{eq:CsFromJumps} are for $s=2$ given by:
	\begin{align}
		J_0 =& \frac{8 i \pi  m M (3 M-2 r_0)}{(l-1) l (l+1) (l+2) r_0^{5/2} (r_0-3 M)^{3/2}}\partial_\theta Y_{lm}^*(\pi/2,0),\\&\nonumber\\
		%J_1 &= \frac{8 i \pi  m M \left(-2 \left(l^2+l-3\right) r_0^3-2 M^2 r_0 \left(6 l (l+1)+3 m^2-22\right)+M r_0^2 \left(10 l (l+1)+2 m^2-31\right)-12 M^3\right)}{(l-1) l (l+1) (l+2) r_0^{7/2} (r_0-3 M)^{3/2} (r_0-2 M)^2}\partial_\theta Y_{lm}^*(\pi/2,0)\\
		J_1 =& \frac{8 i \pi  m M}{(l-1) l (l+1) (l+2) r_0^{7/2} (r_0-3 M)^{3/2} (r_0-2 M)^2} \left\{-2 \left(l^2+l-3\right) r_0^3-2 M^2 r_0 \left(6 l (l+1)+3 m^2-22\right) \right. \nonumber \\ 
		     &\left. +M r_0^2 \left(10 l (l+1)+2 m^2-31\right)-12 M^3\right\}\partial_\theta Y_{lm}^*(\pi/2,0),\\&\nonumber\\
		%J_2 =& 	-\frac{8 i \pi  m M \left(2 M^3 r_0 \left(18 l (l+1)+27 m^2+16\right)-M^2 r_0^2 \left(40 l (l+1)+41 m^2+4\right)+M r_0^3 \left(15 l (l+1)+8 m^2\right)-2 l (l+1) r_0^4-48 M^4\right)}{(l-1) l (l+1) (l+2) r_0^{9/2} (2 M-r_0)^3 (r_0-3 M)^{3/2}}\partial_\theta Y_{lm}^*(\pi/2,0)			\\
		J_2 =& 	-\frac{8 i \pi  m M}{(l-1) l (l+1) (l+2) r_0^{9/2} (2 M-r_0)^3 (r_0-3 M)^{3/2}} \left\{2 M^3 r_0 \left(18 l (l+1)+27 m^2+16\right) \right. 						\nonumber 	\\
			 & \left. -M^2 r_0^2 \left(40 l (l+1)+41 m^2+4\right)+M r_0^3 \left(15 l (l+1)+8 m^2\right)-2 l (l+1) r_0^4-48 M^4\right\}\partial_\theta Y_{lm}^*(\pi/2,0),		\\&\nonumber\\
		%J_3 =& 	\frac{8 i \pi  m M \left(-2 l (l+1) \left(l^2+l-5\right) r_0^5-24 M^4 r_0 \left(5 l (l+1)+10 m^2-42\right)+M r_0^4 \left(l (l+1) \left(14 l (l+1)+4 m^2-61\right)-12 \left(m^2+4\right)\right)+2 M^3 r_0^2 \left((12 l (l+1)+59) m^2+2 l (l+1) (6 l (l+1)+5)+3 m^4-476\right)+M^2 r_0^3 \left(-20 l (l+1) m^2-2 l (l+1) (16 l (l+1)-51)-2 m^4+21 m^2+356\right)-288 M^5\right)}{(l-1) l (l+1) (l+2) r_0^{11/2} (r_0-3 M)^{3/2} (r_0-2 M)^4}\partial_\theta Y_{lm}^*(\pi/2,0)	\\
		J_3 =& 	\frac{8 i \pi  m M}{(l-1) l (l+1) (l+2) r_0^{11/2} (r_0-3 M)^{3/2} (r_0-2 M)^4} \left\{-2 l (l+1) \left(l^2+l-5\right) r_0^5 \right.  \nonumber \\
			 & \left.-24 M^4 r_0 \left(5 l (l+1)+10 m^2-42\right) + M r_0^4 \left(l (l+1) \left(14 l (l+1)+4 m^2-61\right)-12 \left(m^2+4\right)\right) \right. \nonumber \\
			 & \left.+ 2 M^3 r_0^2 \left((12 l (l+1)+59) m^2+2 l (l+1) (6 l (l+1)+5)+3 m^4-476\right) \right. \nonumber \\
			 & \left. + M^2 r_0^3 \left(-20 l (l+1) m^2-2 l (l+1) (16 l (l+1)-51)-2 m^4+21 m^2+356\right)-288 M^5\right\}\partial_\theta Y_{lm}^*(\pi/2,0).	 
	\end{align}
	
	The coefficient of the Dirac delta function in Eq.~\eqref{psirret} is given by:
	\begin{align}
		c_s^\delta \dpsi_2^\delta(r_0) = \frac{-8im M \pi}{\lambda l(l+1) r_0^{3/2}\sqrt{r_0-3M}}\partial_\theta Y_{lm}^*(\pi/2,0).
	\end{align}
	
	\subsection{Jump conditions for $\chi_1$}\label{apdx:Js_chi1}
	For $\chi_1$ the $\mathbf{J}$ vector in Eq.~\eqref{eq:CsFromJumps6thOrder} has components $\mathbf{J} = \{0, 0, J_2, J_3, J_4, J_5 \}$ where:
	\begin{align}
		J_2 =& \frac{16 \pi  \sqrt{r_0 (r_0-3 M)}}{(r_0-2 M)^2}Y_{lm}^*(\pi/2,0), \\&\nonumber\\
		J_3 =& -\frac{8 \pi   \left(42 M^2-21 M r_0+2 r_0^2\right)}{(2 M-r_0)^3 \sqrt{r_0 (r_0-3 M)}}Y_{lm}^*(\pi/2,0),\\&\nonumber\\
		J_4 =& \frac{32 \pi   \left(M^2 r_0 \left(6 l^2+6 l+3 m^2-2\right)-M r_0^2 \left(5 l^2+5 l+m^2-4\right)+l (l+1) r_0^3-24 M^3\right)}{r_0 \sqrt{r_0 (r_0-3 M)} (r_0-2 M)^4}Y_{lm}^*(\pi/2,0), \\&\nonumber\\
		J_5 =& \frac{16 \pi}{r_0^2 (2 M-r_0)^5 \sqrt{r_0 (r_0-3 M)}} \left\{4 M^3 r_0 \left(21 l^2+21 l+33 m^2-23\right)-4 M^2 r_0^2 \left(9 l^2+9 l+15 m^2+16\right) \right .\nonumber \\
		& \left. -5 M r_0^3 \left(3 l^2+3 l-m^2-8\right)+6 l (l+1) r_0^4-72 M^4\right\} Y_{lm}^*(\pi/2,0).
	\end{align}

	\section{Homogeneous even-sector Lorenz gauge metric perturbations for $l \geq 2$ and $\omega \neq 0$} \label{BEven}
	
	The following expressions are derived in Ref.~ \cite{Berndtson:2007gsc}. The reader is referred there for the inhomogeneous case and for expressions for $l=0,1$ and $\omega = 0$. 
	\begin{align}
	H_0(r) \label{H0even} =&-\frac{\lambda (1+\lambda) M
		(-3 M+(3+\lambda) r) \psi_2'}{3 (i \omega)^2 r^3 (3 M+\lambda r)}+
	\frac{(-M+r) \psi_0}{(2 M-r) r}+\frac{2 \left(-2 M^2+M r+(i \omega)^2 
		r^4\right) (\psi_{0b}+\mtaf)}{(2 M-r) r^4}\\&-\frac{\lambda (1+\lambda)	}{3 (i \omega)^2 (2 M-r) r^4 (3 M+\lambda r)^2}\left[18 M^4+3 (-3+4 \lambda) M^3 r+(i \omega)^2 \lambda^2 r^6-3 \lambda M r^3 \left(1+\lambda-2 (i \omega)^2
	r^2\right)\right.\nn\\&\left.+M^2 \left(6 \lambda^2 r^2+9 (i \omega)^2 r^4\right)\right] \psi_2+\frac{4 i \omega (1+\lambda)\psi_1}{2 M-r}
	+\psi_0'+\frac{2 M  (\psi_{0b}'+M_{2af}')}{r^3}+\frac{4 (1+\lambda) M \psi_1'}{i \omega r^3},\nn\\&\nn&\\
	H_1(r)=&-\frac{\lambda(1+\lambda)\dptw}
	{3 \iom r}-\frac{\left(6 M+4 \lambda M-3 r-2\lambda r-2(\iom)^2 r^3\right)\pz}
	{4 \iom M r^2-2 \iom r^3}+\frac{(2+2 \lambda) \po}{2 M r-r^2}
	+\frac{2 \iom (-M+r)(\pzbf)}{(2 M-r) r^2}\\&-\frac{\lambda (1+\lambda)
		\left(3 M^2+3 \lambda M r-\lambda r^2\right) \ptw}
	{3\iom (2 M-r) r^2 (3 M+\lambda r)}
	\nn+\frac{\dpz}{2 \iom r}
	+\frac{2 \iom (\dpzbf)}{r}+\frac{4 (1+\lambda) \dpo}{r},\nn\\&\nn\\
			H_2(r)=&\frac{\lambda (1+\lambda)
		\left(-9 M^2+(3-5 \lambda) M r+2 \lambda r^2\right) \dptw}
	{3 (\iom)^2 r^3 (3 M+\lambda r)}+\frac{(-3 M+2 r) \pz}{(2 M-r) r}+\dpz-
	\frac{4 (1+\lambda) (M-r) \dpo}{\iom r^3}
	\\&+\frac{2 \left(6 M^2-(11+4 \lambda) M r+r^2 \left(4+2 \lambda
		+(\iom)^2 r^2\right)\right) (\pzbf)}{(2 M-r) r^4}+\frac{(-6 M+4 r) (\dpzbf)}{r^3}
	\nn
	\\&
	-\frac{\lambda (1+\lambda)}{3 (\iom)^2 (2 M-r) r^4 (3 M+\lambda r)^2}
	[-54 M^4+3 (9-16 \lambda) M^3 r+\lambda^2 r^4 (2+2 \lambda+(\iom)^2 r^2)
	\nn
	\\&
	+9 M^2 r^2(2 \lambda-2 \lambda^2+(\iom)^2 r^2)+\lambda M r^3 
	(3+5 \lambda-4 \lambda^2+6 (\iom)^2 r^2)] \ptw 
	\nn
	\\&
	+\frac{4 (1+\lambda) \left(-4 (1+\lambda) M+r \left(2+2 \lambda+(\iom)^2 r^2\right)\right) \po}{\iom (2 M-r) r^3},
	\nn\\&\nn\\
	K(r)=&\frac{\lambda (1+\lambda) (2 M-r) \dptw}
	{3 (\iom)^2 r^3}+\frac{\pz}{r}+\frac{(-4 M+2 (2+\lambda) r) (\pzbf)}{r^4}
	+\frac{4 (1+\lambda)^2 \po}{\iom r^3}+\frac{2 (1+\lambda) (2 M-r) \dpo}{\iom r^3}\\&-\frac{\lambda (1+\lambda)
		\left(6 M^2+3 \lambda M r+\lambda (1+\lambda) r^2\right) \ptw}
	{3 (\iom)^2 r^4 (3 M+\lambda r)}+\frac{(4 M-2 r) (\dpzbf)}{r^3},\nn\\\nn&\\
	h_{0}(r)=&\frac{\lambda (2 M-r) \dptw}{3 \iom r}
	+\frac{(-2 M+r) \pz}{2 \iom r^2}+\frac{2 \iom (\pzbf)}{r}
	+\frac{4 (1+\lambda) \po}{r}+\left(1-\frac{2 M}{r}\right) \dpo-\frac{(-2 M+r) \dpz}{2 \iom r} \\&-\frac{\lambda \left(6 M^2+3 \lambda M r
		+\lambda (1+\lambda) r^2\right) \ptw}{3 \iom r^2 (3 M+\lambda r)},
	\nn\\ \nn &\\
	h_{1}(r)=&\frac{\lambda ((3+\lambda) M
	+\lambda (2+\lambda) r) \dptw}{3 (\iom)^2 r (3 M+\lambda r)}
-\frac{r \pz}{4 M-2 r}+\frac{4 \pzbf}{r^2}+\frac{\left(8 M\!+\!8 \lambda M\!-\!4 r\!-\!4 \lambda r\!
	+\!(\iom)^2 r^3\right) \po}{2\iom M r^2-\iom r^3}
\\&-\frac{\lambda}
{3 (\iom)^2 (2 M\!-\!r) r^2 (3 M\!+\!\lambda
	r)^2}\left[9 (\iom)^2 M^2 r^3+2 \lambda^3 r^2 (-2 M\!+\!r)
\!+\!\lambda^2 r \left(-12 M^2\!+\!2 M r\!+\!2 r^2\!+\!(\iom)^2 r^4\right)\right.\nn\\&\left.+3 \lambda M \left(-4 M^2+r^2+2 (\iom)^2 r^4\right)\right]\ptw-\frac{2 (\dpzbf)}{r}-\frac{2 (2 M+r+2 \lambda r) \dpo}{\iom r^2},\nn\\\nn&\\
G(r) \label{Geven}=&-\frac{\lambda (3+2 \lambda)(2 M-r)\dptw}
{6 (\iom)^2 r^2 (3 M+\lambda r)}-\frac{\pzbf}{r^3}-\frac{2 (1+\lambda) \po}
{\iom r^3}+\frac{(2 M-r) \dpo}{\iom r^3}\\&+\frac{\left[4 \lambda^3 r^2+\lambda^4 r^2+27 (\iom)^2 M^2 r^2+9 \lambda M \left(M+2 (\iom)^2 r^3\right)+3 \lambda^2 \left(M^2+M
	r+r^2+(\iom)^2 r^4\right)\right]}{6 (\iom)^2 r^3 (3 M+\lambda r)^2}
	\ptw. \nn
	\end{align}
	
	\twocolumngrid
	\newpage
	\addcontentsline{toc}{chapter}{Bibliography}
	\bibliographystyle{apsrev4-2}
	\bibliography{bibfile}

\begin{thebibliography}{59}%
\makeatletter
\providecommand \@ifxundefined [1]{%
 \@ifx{#1\undefined}
}%
\providecommand \@ifnum [1]{%
 \ifnum #1\expandafter \@firstoftwo
 \else \expandafter \@secondoftwo
 \fi
}%
\providecommand \@ifx [1]{%
 \ifx #1\expandafter \@firstoftwo
 \else \expandafter \@secondoftwo
 \fi
}%
\providecommand \natexlab [1]{#1}%
\providecommand \enquote  [1]{``#1''}%
\providecommand \bibnamefont  [1]{#1}%
\providecommand \bibfnamefont [1]{#1}%
\providecommand \citenamefont [1]{#1}%
\providecommand \href@noop [0]{\@secondoftwo}%
\providecommand \href [0]{\begingroup \@sanitize@url \@href}%
\providecommand \@href[1]{\@@startlink{#1}\@@href}%
\providecommand \@@href[1]{\endgroup#1\@@endlink}%
\providecommand \@sanitize@url [0]{\catcode `\\12\catcode `\$12\catcode
  `\&12\catcode `\#12\catcode `\^12\catcode `\_12\catcode `\%12\relax}%
\providecommand \@@startlink[1]{}%
\providecommand \@@endlink[0]{}%
\providecommand \url  [0]{\begingroup\@sanitize@url \@url }%
\providecommand \@url [1]{\endgroup\@href {#1}{\urlprefix }}%
\providecommand \urlprefix  [0]{URL }%
\providecommand \Eprint [0]{\href }%
\providecommand \doibase [0]{https://doi.org/}%
\providecommand \selectlanguage [0]{\@gobble}%
\providecommand \bibinfo  [0]{\@secondoftwo}%
\providecommand \bibfield  [0]{\@secondoftwo}%
\providecommand \translation [1]{[#1]}%
\providecommand \BibitemOpen [0]{}%
\providecommand \bibitemStop [0]{}%
\providecommand \bibitemNoStop [0]{.\EOS\space}%
\providecommand \EOS [0]{\spacefactor3000\relax}%
\providecommand \BibitemShut  [1]{\csname bibitem#1\endcsname}%
\let\auto@bib@innerbib\@empty
%</preamble>
\bibitem [{\citenamefont {Abbott}\ \emph {et~al.}(2016)\citenamefont {Abbott}
  \emph {et~al.}}]{PhysRevLett.116.061102}%
  \BibitemOpen
  \bibfield  {author} {\bibinfo {author} {\bibfnamefont {R.}~\bibnamefont
  {Abbott}} \emph {et~al.} (\bibinfo {collaboration} {LIGO Scientific
  Collaboration and Virgo Collaboration}),\ }\href
  {https://doi.org/10.1103/PhysRevLett.116.061102} {\bibfield  {journal}
  {\bibinfo  {journal} {Phys. Rev. Lett.}\ }\textbf {\bibinfo {volume} {116}},\
  \bibinfo {pages} {061102} (\bibinfo {year} {2016})}\BibitemShut {NoStop}%
\bibitem [{\citenamefont {Abbott}\ \emph
  {et~al.}(2021{\natexlab{a}})\citenamefont {Abbott} \emph
  {et~al.}}]{LIGOScientific:2021djp}%
  \BibitemOpen
  \bibfield  {author} {\bibinfo {author} {\bibfnamefont {R.}~\bibnamefont
  {Abbott}} \emph {et~al.} (\bibinfo {collaboration} {LIGO Scientific, VIRGO,
  KAGRA}),\ }\href@noop {} {\  (\bibinfo {year} {2021}{\natexlab{a}})},\
  \Eprint {https://arxiv.org/abs/2111.03606} {arXiv:2111.03606 [gr-qc]}
  \BibitemShut {NoStop}%
\bibitem [{\citenamefont {Abbott}\ \emph {et~al.}(2017)\citenamefont {Abbott}
  \emph {et~al.}}]{PhysRevLett.119.161101}%
  \BibitemOpen
  \bibfield  {author} {\bibinfo {author} {\bibfnamefont {R.}~\bibnamefont
  {Abbott}} \emph {et~al.} (\bibinfo {collaboration} {LIGO Scientific
  Collaboration and Virgo Collaboration}),\ }\href
  {https://doi.org/10.1103/PhysRevLett.119.161101} {\bibfield  {journal}
  {\bibinfo  {journal} {Phys. Rev. Lett.}\ }\textbf {\bibinfo {volume} {119}},\
  \bibinfo {pages} {161101} (\bibinfo {year} {2017})}\BibitemShut {NoStop}%
\bibitem [{\citenamefont {Abbott}\ \emph
  {et~al.}(2021{\natexlab{b}})\citenamefont {Abbott} \emph
  {et~al.}}]{LIGOScientific:2021qlt}%
  \BibitemOpen
  \bibfield  {author} {\bibinfo {author} {\bibfnamefont {R.}~\bibnamefont
  {Abbott}} \emph {et~al.} (\bibinfo {collaboration} {LIGO Scientific
  Collaboration, KAGRA and Virgo Collaboration}),\ }\href
  {https://doi.org/10.3847/2041-8213/ac082e} {\bibfield  {journal} {\bibinfo
  {journal} {Astrophys. J. Lett.}\ }\textbf {\bibinfo {volume} {915}},\
  \bibinfo {pages} {L5} (\bibinfo {year} {2021}{\natexlab{b}})},\ \Eprint
  {https://arxiv.org/abs/2106.15163} {arXiv:2106.15163 [astro-ph.HE]}
  \BibitemShut {NoStop}%
\bibitem [{\citenamefont {Amaro-Seoane}\ \emph {et~al.}(2017)\citenamefont
  {Amaro-Seoane} \emph {et~al.}}]{LISA:2017pwj}%
  \BibitemOpen
  \bibfield  {author} {\bibinfo {author} {\bibfnamefont {P.}~\bibnamefont
  {Amaro-Seoane}} \emph {et~al.} (\bibinfo {collaboration} {LISA}),\
  }\href@noop {} {\  (\bibinfo {year} {2017})},\ \Eprint
  {https://arxiv.org/abs/1702.00786} {arXiv:1702.00786 [astro-ph.IM]}
  \BibitemShut {NoStop}%
\bibitem [{\citenamefont {Berry}\ \emph {et~al.}(2019)\citenamefont {Berry},
  \citenamefont {Hughes}, \citenamefont {Sopuerta}, \citenamefont {Chua},
  \citenamefont {Heffernan}, \citenamefont {Holley-Bockelmann}, \citenamefont
  {Mihaylov}, \citenamefont {Miller},\ and\ \citenamefont
  {Sesana}}]{Berry:2019wgg}%
  \BibitemOpen
  \bibfield  {author} {\bibinfo {author} {\bibfnamefont {C.~P.~L.}\
  \bibnamefont {Berry}}, \bibinfo {author} {\bibfnamefont {S.~A.}\ \bibnamefont
  {Hughes}}, \bibinfo {author} {\bibfnamefont {C.~F.}\ \bibnamefont
  {Sopuerta}}, \bibinfo {author} {\bibfnamefont {A.~J.~K.}\ \bibnamefont
  {Chua}}, \bibinfo {author} {\bibfnamefont {A.}~\bibnamefont {Heffernan}},
  \bibinfo {author} {\bibfnamefont {K.}~\bibnamefont {Holley-Bockelmann}},
  \bibinfo {author} {\bibfnamefont {D.~P.}\ \bibnamefont {Mihaylov}}, \bibinfo
  {author} {\bibfnamefont {M.~C.}\ \bibnamefont {Miller}},\ and\ \bibinfo
  {author} {\bibfnamefont {A.}~\bibnamefont {Sesana}},\ }\href@noop {} {\
  (\bibinfo {year} {2019})},\ \Eprint {https://arxiv.org/abs/1903.03686}
  {arXiv:1903.03686 [astro-ph.HE]} \BibitemShut {NoStop}%
\bibitem [{\citenamefont {Babak}\ \emph {et~al.}(2017)\citenamefont {Babak},
  \citenamefont {Gair}, \citenamefont {Sesana}, \citenamefont {Barausse},
  \citenamefont {Sopuerta}, \citenamefont {Berry}, \citenamefont {Berti},
  \citenamefont {Amaro-Seoane}, \citenamefont {Petiteau},\ and\ \citenamefont
  {Klein}}]{Babak:2017tow}%
  \BibitemOpen
  \bibfield  {author} {\bibinfo {author} {\bibfnamefont {S.}~\bibnamefont
  {Babak}}, \bibinfo {author} {\bibfnamefont {J.}~\bibnamefont {Gair}},
  \bibinfo {author} {\bibfnamefont {A.}~\bibnamefont {Sesana}}, \bibinfo
  {author} {\bibfnamefont {E.}~\bibnamefont {Barausse}}, \bibinfo {author}
  {\bibfnamefont {C.~F.}\ \bibnamefont {Sopuerta}}, \bibinfo {author}
  {\bibfnamefont {C.~P.~L.}\ \bibnamefont {Berry}}, \bibinfo {author}
  {\bibfnamefont {E.}~\bibnamefont {Berti}}, \bibinfo {author} {\bibfnamefont
  {P.}~\bibnamefont {Amaro-Seoane}}, \bibinfo {author} {\bibfnamefont
  {A.}~\bibnamefont {Petiteau}},\ and\ \bibinfo {author} {\bibfnamefont
  {A.}~\bibnamefont {Klein}},\ }\href
  {https://doi.org/10.1103/PhysRevD.95.103012} {\bibfield  {journal} {\bibinfo
  {journal} {Phys. Rev. D}\ }\textbf {\bibinfo {volume} {95}},\ \bibinfo
  {pages} {103012} (\bibinfo {year} {2017})},\ \Eprint
  {https://arxiv.org/abs/1703.09722} {arXiv:1703.09722 [gr-qc]} \BibitemShut
  {NoStop}%
\bibitem [{\citenamefont {Miller}\ and\ \citenamefont
  {Pound}(2021)}]{Miller:2020bft}%
  \BibitemOpen
  \bibfield  {author} {\bibinfo {author} {\bibfnamefont {J.}~\bibnamefont
  {Miller}}\ and\ \bibinfo {author} {\bibfnamefont {A.}~\bibnamefont {Pound}},\
  }\href {https://doi.org/10.1103/PhysRevD.103.064048} {\bibfield  {journal}
  {\bibinfo  {journal} {Phys. Rev. D}\ }\textbf {\bibinfo {volume} {103}},\
  \bibinfo {pages} {064048} (\bibinfo {year} {2021})},\ \Eprint
  {https://arxiv.org/abs/2006.11263} {arXiv:2006.11263 [gr-qc]} \BibitemShut
  {NoStop}%
\bibitem [{\citenamefont {Hinderer}\ and\ \citenamefont
  {Flanagan}(2008)}]{Hinderer:2008dm}%
  \BibitemOpen
  \bibfield  {author} {\bibinfo {author} {\bibfnamefont {T.}~\bibnamefont
  {Hinderer}}\ and\ \bibinfo {author} {\bibfnamefont {E.~E.}\ \bibnamefont
  {Flanagan}},\ }\href {https://doi.org/10.1103/PhysRevD.78.064028} {\bibfield
  {journal} {\bibinfo  {journal} {Phys. Rev. D}\ }\textbf {\bibinfo {volume}
  {78}},\ \bibinfo {pages} {064028} (\bibinfo {year} {2008})},\ \Eprint
  {https://arxiv.org/abs/0805.3337} {arXiv:0805.3337 [gr-qc]} \BibitemShut
  {NoStop}%
\bibitem [{\citenamefont {Poisson}\ \emph {et~al.}(2011)\citenamefont
  {Poisson}, \citenamefont {Pound},\ and\ \citenamefont
  {Vega}}]{Poisson:2011nh}%
  \BibitemOpen
  \bibfield  {author} {\bibinfo {author} {\bibfnamefont {E.}~\bibnamefont
  {Poisson}}, \bibinfo {author} {\bibfnamefont {A.}~\bibnamefont {Pound}},\
  and\ \bibinfo {author} {\bibfnamefont {I.}~\bibnamefont {Vega}},\ }\href
  {https://doi.org/10.12942/lrr-2011-7} {\bibfield  {journal} {\bibinfo
  {journal} {Living Rev. Rel.}\ }\textbf {\bibinfo {volume} {14}},\ \bibinfo
  {pages} {7} (\bibinfo {year} {2011})},\ \Eprint
  {https://arxiv.org/abs/1102.0529} {arXiv:1102.0529 [gr-qc]} \BibitemShut
  {NoStop}%
\bibitem [{\citenamefont {Barack}\ and\ \citenamefont
  {Pound}(2019)}]{Barack:2018yvs}%
  \BibitemOpen
  \bibfield  {author} {\bibinfo {author} {\bibfnamefont {L.}~\bibnamefont
  {Barack}}\ and\ \bibinfo {author} {\bibfnamefont {A.}~\bibnamefont {Pound}},\
  }\href {https://doi.org/10.1088/1361-6633/aae552} {\bibfield  {journal}
  {\bibinfo  {journal} {Rept. Prog. Phys.}\ }\textbf {\bibinfo {volume} {82}},\
  \bibinfo {pages} {016904} (\bibinfo {year} {2019})},\ \Eprint
  {https://arxiv.org/abs/1805.10385} {arXiv:1805.10385 [gr-qc]} \BibitemShut
  {NoStop}%
\bibitem [{\citenamefont {Pound}(2012{\natexlab{a}})}]{Pound:2012nt}%
  \BibitemOpen
  \bibfield  {author} {\bibinfo {author} {\bibfnamefont {A.}~\bibnamefont
  {Pound}},\ }\href {https://doi.org/10.1103/PhysRevLett.109.051101} {\bibfield
   {journal} {\bibinfo  {journal} {Phys. Rev. Lett.}\ }\textbf {\bibinfo
  {volume} {109}},\ \bibinfo {pages} {051101} (\bibinfo {year}
  {2012}{\natexlab{a}})},\ \Eprint {https://arxiv.org/abs/1201.5089}
  {arXiv:1201.5089 [gr-qc]} \BibitemShut {NoStop}%
\bibitem [{\citenamefont {Pound}\ and\ \citenamefont
  {Miller}(2014)}]{Pound:2014xva}%
  \BibitemOpen
  \bibfield  {author} {\bibinfo {author} {\bibfnamefont {A.}~\bibnamefont
  {Pound}}\ and\ \bibinfo {author} {\bibfnamefont {J.}~\bibnamefont {Miller}},\
  }\href {https://doi.org/10.1103/PhysRevD.89.104020} {\bibfield  {journal}
  {\bibinfo  {journal} {Phys. Rev. D}\ }\textbf {\bibinfo {volume} {89}},\
  \bibinfo {pages} {104020} (\bibinfo {year} {2014})},\ \Eprint
  {https://arxiv.org/abs/1403.1843} {arXiv:1403.1843 [gr-qc]} \BibitemShut
  {NoStop}%
\bibitem [{\citenamefont {Barack}\ and\ \citenamefont
  {Lousto}(2005)}]{Barack:2005nr}%
  \BibitemOpen
  \bibfield  {author} {\bibinfo {author} {\bibfnamefont {L.}~\bibnamefont
  {Barack}}\ and\ \bibinfo {author} {\bibfnamefont {C.~O.}\ \bibnamefont
  {Lousto}},\ }\href {https://doi.org/10.1103/PhysRevD.72.104026} {\bibfield
  {journal} {\bibinfo  {journal} {Phys. Rev. D}\ }\textbf {\bibinfo {volume}
  {72}},\ \bibinfo {pages} {104026} (\bibinfo {year} {2005})},\ \Eprint
  {https://arxiv.org/abs/gr-qc/0510019} {arXiv:gr-qc/0510019} \BibitemShut
  {NoStop}%
\bibitem [{\citenamefont {Barack}\ and\ \citenamefont
  {Sago}(2007)}]{Barack:2007tm}%
  \BibitemOpen
  \bibfield  {author} {\bibinfo {author} {\bibfnamefont {L.}~\bibnamefont
  {Barack}}\ and\ \bibinfo {author} {\bibfnamefont {N.}~\bibnamefont {Sago}},\
  }\href {https://doi.org/10.1103/PhysRevD.75.064021} {\bibfield  {journal}
  {\bibinfo  {journal} {Phys. Rev. D}\ }\textbf {\bibinfo {volume} {75}},\
  \bibinfo {pages} {064021} (\bibinfo {year} {2007})},\ \Eprint
  {https://arxiv.org/abs/gr-qc/0701069} {arXiv:gr-qc/0701069} \BibitemShut
  {NoStop}%
\bibitem [{\citenamefont {Akcay}(2011)}]{Akcay:2010dx}%
  \BibitemOpen
  \bibfield  {author} {\bibinfo {author} {\bibfnamefont {S.}~\bibnamefont
  {Akcay}},\ }\href {https://doi.org/10.1103/PhysRevD.83.124026} {\bibfield
  {journal} {\bibinfo  {journal} {Phys. Rev. D}\ }\textbf {\bibinfo {volume}
  {83}},\ \bibinfo {pages} {124026} (\bibinfo {year} {2011})},\ \Eprint
  {https://arxiv.org/abs/1012.5860} {arXiv:1012.5860 [gr-qc]} \BibitemShut
  {NoStop}%
\bibitem [{\citenamefont {Akcay}\ \emph {et~al.}(2013)\citenamefont {Akcay},
  \citenamefont {Warburton},\ and\ \citenamefont {Barack}}]{Akcay:2013wfa}%
  \BibitemOpen
  \bibfield  {author} {\bibinfo {author} {\bibfnamefont {S.}~\bibnamefont
  {Akcay}}, \bibinfo {author} {\bibfnamefont {N.}~\bibnamefont {Warburton}},\
  and\ \bibinfo {author} {\bibfnamefont {L.}~\bibnamefont {Barack}},\ }\href
  {https://doi.org/10.1103/PhysRevD.88.104009} {\bibfield  {journal} {\bibinfo
  {journal} {Phys. Rev. D}\ }\textbf {\bibinfo {volume} {88}},\ \bibinfo
  {pages} {104009} (\bibinfo {year} {2013})},\ \Eprint
  {https://arxiv.org/abs/1308.5223} {arXiv:1308.5223 [gr-qc]} \BibitemShut
  {NoStop}%
\bibitem [{\citenamefont {Wardell}\ and\ \citenamefont
  {Warburton}(2015)}]{Wardell:2015ada}%
  \BibitemOpen
  \bibfield  {author} {\bibinfo {author} {\bibfnamefont {B.}~\bibnamefont
  {Wardell}}\ and\ \bibinfo {author} {\bibfnamefont {N.}~\bibnamefont
  {Warburton}},\ }\href {https://doi.org/10.1103/PhysRevD.92.084019} {\bibfield
   {journal} {\bibinfo  {journal} {Phys. Rev. D}\ }\textbf {\bibinfo {volume}
  {92}},\ \bibinfo {pages} {084019} (\bibinfo {year} {2015})},\ \Eprint
  {https://arxiv.org/abs/1505.07841} {arXiv:1505.07841 [gr-qc]} \BibitemShut
  {NoStop}%
\bibitem [{\citenamefont {Miller}\ \emph {et~al.}()\citenamefont {Miller},
  \citenamefont {Leather}, \citenamefont {Pound},\ and\ \citenamefont
  {Warburton}}]{Miller_etal}%
  \BibitemOpen
  \bibinfo {author} {\bibfnamefont {J.}~\bibnamefont {Miller}}, \bibinfo
  {author} {\bibfnamefont {B.}~\bibnamefont {Leather}}, \bibinfo {author}
  {\bibfnamefont {A.}~\bibnamefont {Pound}},\ and\ \bibinfo {author}
  {\bibfnamefont {N.}~\bibnamefont {Warburton}}\BibitemShut {NoStop}%
\bibitem [{\citenamefont {Berndtson}(2007)}]{Berndtson:2007gsc}%
  \BibitemOpen
\bibfield  {author} {  }\bibfield  {author} {\bibinfo {author} {\bibfnamefont
  {M.~V.}\ \bibnamefont {Berndtson}},\ }\emph {\bibinfo {title} {{Harmonic
  gauge perturbations of the Schwarzschild metric}}},\ \href@noop {} {Ph.D.
  thesis},\ \bibinfo  {school} {University of Colorado} (\bibinfo {year}
  {2007}),\ \Eprint {https://arxiv.org/abs/0904.0033} {arXiv:0904.0033 [gr-qc]}
  \BibitemShut {NoStop}%
\bibitem [{\citenamefont {Hopper}\ and\ \citenamefont
  {Evans}(2013)}]{Hopper:2012ty}%
  \BibitemOpen
  \bibfield  {author} {\bibinfo {author} {\bibfnamefont {S.}~\bibnamefont
  {Hopper}}\ and\ \bibinfo {author} {\bibfnamefont {C.~R.}\ \bibnamefont
  {Evans}},\ }\href {https://doi.org/10.1103/PhysRevD.87.064008} {\bibfield
  {journal} {\bibinfo  {journal} {Phys. Rev. D}\ }\textbf {\bibinfo {volume}
  {87}},\ \bibinfo {pages} {064008} (\bibinfo {year} {2013})},\ \Eprint
  {https://arxiv.org/abs/1210.7969} {arXiv:1210.7969 [gr-qc]} \BibitemShut
  {NoStop}%
\bibitem [{\citenamefont {Osburn}\ \emph {et~al.}(2014)\citenamefont {Osburn},
  \citenamefont {Forseth}, \citenamefont {Evans},\ and\ \citenamefont
  {Hopper}}]{Osburn:2014hoa}%
  \BibitemOpen
  \bibfield  {author} {\bibinfo {author} {\bibfnamefont {T.}~\bibnamefont
  {Osburn}}, \bibinfo {author} {\bibfnamefont {E.}~\bibnamefont {Forseth}},
  \bibinfo {author} {\bibfnamefont {C.~R.}\ \bibnamefont {Evans}},\ and\
  \bibinfo {author} {\bibfnamefont {S.}~\bibnamefont {Hopper}},\ }\href
  {https://doi.org/10.1103/PhysRevD.90.104031} {\bibfield  {journal} {\bibinfo
  {journal} {Phys. Rev. D}\ }\textbf {\bibinfo {volume} {90}},\ \bibinfo
  {pages} {104031} (\bibinfo {year} {2014})},\ \Eprint
  {https://arxiv.org/abs/1409.4419} {arXiv:1409.4419 [gr-qc]} \BibitemShut
  {NoStop}%
\bibitem [{BHP()}]{BHPToolkit}%
  \BibitemOpen
  \href@noop {} {\bibinfo {title} {Black hole perturbation toolkit}},\ \bibinfo
  {howpublished} {bhptoolkit.org}\BibitemShut {NoStop}%
\bibitem [{\citenamefont {Dolan}\ \emph {et~al.}(2022)\citenamefont {Dolan},
  \citenamefont {Kavanagh},\ and\ \citenamefont {Wardell}}]{Dolan:2021ijg}%
  \BibitemOpen
  \bibfield  {author} {\bibinfo {author} {\bibfnamefont {S.~R.}\ \bibnamefont
  {Dolan}}, \bibinfo {author} {\bibfnamefont {C.}~\bibnamefont {Kavanagh}},\
  and\ \bibinfo {author} {\bibfnamefont {B.}~\bibnamefont {Wardell}},\ }\href
  {https://doi.org/10.1103/PhysRevLett.128.151101} {\bibfield  {journal}
  {\bibinfo  {journal} {Phys. Rev. Lett.}\ }\textbf {\bibinfo {volume} {128}},\
  \bibinfo {pages} {151101} (\bibinfo {year} {2022})},\ \Eprint
  {https://arxiv.org/abs/2108.06344} {arXiv:2108.06344 [gr-qc]} \BibitemShut
  {NoStop}%
\bibitem [{\citenamefont {Durkan}\ and\ \citenamefont {Dolan}(2021)}]{me}%
  \BibitemOpen
  \bibfield  {author} {\bibinfo {author} {\bibfnamefont {L.}~\bibnamefont
  {Durkan}}\ and\ \bibinfo {author} {\bibfnamefont {S.}~\bibnamefont {Dolan}},\
  }\href@noop {} {\bibinfo {title} {{Private Communication}}} (\bibinfo {year}
  {2021})\BibitemShut {NoStop}%
\bibitem [{\citenamefont {Barack}\ and\ \citenamefont
  {Sago}(2010)}]{Barack:2010tm}%
  \BibitemOpen
  \bibfield  {author} {\bibinfo {author} {\bibfnamefont {L.}~\bibnamefont
  {Barack}}\ and\ \bibinfo {author} {\bibfnamefont {N.}~\bibnamefont {Sago}},\
  }\href {https://doi.org/10.1103/PhysRevD.81.084021} {\bibfield  {journal}
  {\bibinfo  {journal} {Phys. Rev. D}\ }\textbf {\bibinfo {volume} {81}},\
  \bibinfo {pages} {084021} (\bibinfo {year} {2010})},\ \Eprint
  {https://arxiv.org/abs/1002.2386} {arXiv:1002.2386 [gr-qc]} \BibitemShut
  {NoStop}%
\bibitem [{\citenamefont {Pound}\ \emph {et~al.}(2020)\citenamefont {Pound},
  \citenamefont {Wardell}, \citenamefont {Warburton},\ and\ \citenamefont
  {Miller}}]{Pound:2019lzj}%
  \BibitemOpen
  \bibfield  {author} {\bibinfo {author} {\bibfnamefont {A.}~\bibnamefont
  {Pound}}, \bibinfo {author} {\bibfnamefont {B.}~\bibnamefont {Wardell}},
  \bibinfo {author} {\bibfnamefont {N.}~\bibnamefont {Warburton}},\ and\
  \bibinfo {author} {\bibfnamefont {J.}~\bibnamefont {Miller}},\ }\href
  {https://doi.org/10.1103/PhysRevLett.124.021101} {\bibfield  {journal}
  {\bibinfo  {journal} {Phys. Rev. Lett.}\ }\textbf {\bibinfo {volume} {124}},\
  \bibinfo {pages} {021101} (\bibinfo {year} {2020})},\ \Eprint
  {https://arxiv.org/abs/1908.07419} {arXiv:1908.07419 [gr-qc]} \BibitemShut
  {NoStop}%
\bibitem [{\citenamefont {Warburton}\ \emph {et~al.}(2021)\citenamefont
  {Warburton}, \citenamefont {Pound}, \citenamefont {Wardell}, \citenamefont
  {Miller},\ and\ \citenamefont {Durkan}}]{Warburton:2021kwk}%
  \BibitemOpen
  \bibfield  {author} {\bibinfo {author} {\bibfnamefont {N.}~\bibnamefont
  {Warburton}}, \bibinfo {author} {\bibfnamefont {A.}~\bibnamefont {Pound}},
  \bibinfo {author} {\bibfnamefont {B.}~\bibnamefont {Wardell}}, \bibinfo
  {author} {\bibfnamefont {J.}~\bibnamefont {Miller}},\ and\ \bibinfo {author}
  {\bibfnamefont {L.}~\bibnamefont {Durkan}},\ }\href
  {https://doi.org/10.1103/PhysRevLett.127.151102} {\bibfield  {journal}
  {\bibinfo  {journal} {Phys. Rev. Lett.}\ }\textbf {\bibinfo {volume} {127}},\
  \bibinfo {pages} {151102} (\bibinfo {year} {2021})},\ \Eprint
  {https://arxiv.org/abs/2107.01298} {arXiv:2107.01298 [gr-qc]} \BibitemShut
  {NoStop}%
\bibitem [{\citenamefont {Wardell}\ \emph {et~al.}(2021)\citenamefont
  {Wardell}, \citenamefont {Pound}, \citenamefont {Warburton}, \citenamefont
  {Miller}, \citenamefont {Durkan},\ and\ \citenamefont
  {Le~Tiec}}]{Wardell:2021fyy}%
  \BibitemOpen
  \bibfield  {author} {\bibinfo {author} {\bibfnamefont {B.}~\bibnamefont
  {Wardell}}, \bibinfo {author} {\bibfnamefont {A.}~\bibnamefont {Pound}},
  \bibinfo {author} {\bibfnamefont {N.}~\bibnamefont {Warburton}}, \bibinfo
  {author} {\bibfnamefont {J.}~\bibnamefont {Miller}}, \bibinfo {author}
  {\bibfnamefont {L.}~\bibnamefont {Durkan}},\ and\ \bibinfo {author}
  {\bibfnamefont {A.}~\bibnamefont {Le~Tiec}},\ }\href@noop {} {\  (\bibinfo
  {year} {2021})},\ \Eprint {https://arxiv.org/abs/2112.12265}
  {arXiv:2112.12265 [gr-qc]} \BibitemShut {NoStop}%
\bibitem [{\citenamefont {Regge}\ and\ \citenamefont
  {Wheeler}(1957)}]{PhysRev.108.1063}%
  \BibitemOpen
  \bibfield  {author} {\bibinfo {author} {\bibfnamefont {T.}~\bibnamefont
  {Regge}}\ and\ \bibinfo {author} {\bibfnamefont {J.~A.}\ \bibnamefont
  {Wheeler}},\ }\href {https://doi.org/10.1103/PhysRev.108.1063} {\bibfield
  {journal} {\bibinfo  {journal} {Phys. Rev.}\ }\textbf {\bibinfo {volume}
  {108}},\ \bibinfo {pages} {1063} (\bibinfo {year} {1957})}\BibitemShut
  {NoStop}%
\bibitem [{\citenamefont {Maggiore}(2018)}]{Maggiore}%
  \BibitemOpen
  \bibfield  {author} {\bibinfo {author} {\bibfnamefont {M.}~\bibnamefont
  {Maggiore}},\ }\href@noop {} {\emph {\bibinfo {title} {{Gravitational Waves:
  Volume 2: Astrophysics and Cosmology.}}}}\ (\bibinfo  {publisher} {Oxford
  University Press},\ \bibinfo {year} {2018})\BibitemShut {NoStop}%
\bibitem [{\citenamefont {Pound}(2012{\natexlab{b}})}]{Pound:2012dk}%
  \BibitemOpen
  \bibfield  {author} {\bibinfo {author} {\bibfnamefont {A.}~\bibnamefont
  {Pound}},\ }\href {https://doi.org/10.1103/PhysRevD.86.084019} {\bibfield
  {journal} {\bibinfo  {journal} {Phys. Rev. D}\ }\textbf {\bibinfo {volume}
  {86}},\ \bibinfo {pages} {084019} (\bibinfo {year} {2012}{\natexlab{b}})},\
  \Eprint {https://arxiv.org/abs/1206.6538} {arXiv:1206.6538 [gr-qc]}
  \BibitemShut {NoStop}%
\bibitem [{\citenamefont {Gralla}(2012)}]{Gralla:2012db}%
  \BibitemOpen
  \bibfield  {author} {\bibinfo {author} {\bibfnamefont {S.~E.}\ \bibnamefont
  {Gralla}},\ }\href {https://doi.org/10.1103/PhysRevD.85.124011} {\bibfield
  {journal} {\bibinfo  {journal} {Phys. Rev. D}\ }\textbf {\bibinfo {volume}
  {85}},\ \bibinfo {pages} {124011} (\bibinfo {year} {2012})},\ \Eprint
  {https://arxiv.org/abs/1203.3189} {arXiv:1203.3189 [gr-qc]} \BibitemShut
  {NoStop}%
\bibitem [{\citenamefont {Miller}\ \emph {et~al.}(2016)\citenamefont {Miller},
  \citenamefont {Wardell},\ and\ \citenamefont {Pound}}]{Miller:2016hjv}%
  \BibitemOpen
  \bibfield  {author} {\bibinfo {author} {\bibfnamefont {J.}~\bibnamefont
  {Miller}}, \bibinfo {author} {\bibfnamefont {B.}~\bibnamefont {Wardell}},\
  and\ \bibinfo {author} {\bibfnamefont {A.}~\bibnamefont {Pound}},\ }\href
  {https://doi.org/10.1103/PhysRevD.94.104018} {\bibfield  {journal} {\bibinfo
  {journal} {Phys. Rev. D}\ }\textbf {\bibinfo {volume} {94}},\ \bibinfo
  {pages} {104018} (\bibinfo {year} {2016})},\ \Eprint
  {https://arxiv.org/abs/1608.06783} {arXiv:1608.06783 [gr-qc]} \BibitemShut
  {NoStop}%
\bibitem [{\citenamefont {Panosso~Macedo}\ \emph {et~al.}(2022)\citenamefont
  {Panosso~Macedo}, \citenamefont {Leather}, \citenamefont {Warburton},
  \citenamefont {Wardell},\ and\ \citenamefont
  {Zengino\u{g}lu}}]{PanossoMacedo:2022fdi}%
  \BibitemOpen
  \bibfield  {author} {\bibinfo {author} {\bibfnamefont {R.}~\bibnamefont
  {Panosso~Macedo}}, \bibinfo {author} {\bibfnamefont {B.}~\bibnamefont
  {Leather}}, \bibinfo {author} {\bibfnamefont {N.}~\bibnamefont {Warburton}},
  \bibinfo {author} {\bibfnamefont {B.}~\bibnamefont {Wardell}},\ and\ \bibinfo
  {author} {\bibfnamefont {A.}~\bibnamefont {Zengino\u{g}lu}},\ }\href
  {https://doi.org/10.1103/PhysRevD.105.104033} {\bibfield  {journal} {\bibinfo
   {journal} {Phys. Rev. D}\ }\textbf {\bibinfo {volume} {105}},\ \bibinfo
  {pages} {104033} (\bibinfo {year} {2022})},\ \Eprint
  {https://arxiv.org/abs/2202.01794} {arXiv:2202.01794 [gr-qc]} \BibitemShut
  {NoStop}%
\bibitem [{\citenamefont {Mathews}\ \emph {et~al.}(2022)\citenamefont
  {Mathews}, \citenamefont {Pound},\ and\ \citenamefont
  {Wardell}}]{Mathews:2021rod}%
  \BibitemOpen
  \bibfield  {author} {\bibinfo {author} {\bibfnamefont {J.}~\bibnamefont
  {Mathews}}, \bibinfo {author} {\bibfnamefont {A.}~\bibnamefont {Pound}},\
  and\ \bibinfo {author} {\bibfnamefont {B.}~\bibnamefont {Wardell}},\ }\href
  {https://doi.org/10.1103/PhysRevD.105.084031} {\bibfield  {journal} {\bibinfo
   {journal} {Phys. Rev. D}\ }\textbf {\bibinfo {volume} {105}},\ \bibinfo
  {pages} {084031} (\bibinfo {year} {2022})},\ \Eprint
  {https://arxiv.org/abs/2112.13069} {arXiv:2112.13069 [gr-qc]} \BibitemShut
  {NoStop}%
\bibitem [{\citenamefont {Trench}(2013)}]{ODEs}%
  \BibitemOpen
  \bibfield  {author} {\bibinfo {author} {\bibfnamefont {W.~F.}\ \bibnamefont
  {Trench}},\ }\href@noop {} {\emph {\bibinfo {title} {{Book: Elementary
  Differential Equations with Boundary Value Problems (Trench)}}}}\ (\bibinfo
  {publisher} {Brooks/Cole Thompson Learning},\ \bibinfo {year}
  {2013})\BibitemShut {NoStop}%
\bibitem [{\citenamefont {Durkan}\ and\ \citenamefont
  {Warburton}(2022)}]{suppmat}%
  \BibitemOpen
  \bibfield  {author} {\bibinfo {author} {\bibfnamefont {L.}~\bibnamefont
  {Durkan}}\ and\ \bibinfo {author} {\bibfnamefont {N.}~\bibnamefont
  {Warburton}},\ }\href@noop {} {\bibinfo {title} {{Supplementary Material for
  Slow evolution of the metric perturbation due to a quasicircular inspiral
  into a Schwarzschild black hole}}} (\bibinfo {year} {2022}),\ \bibinfo {note}
  {boundary conditions for higher order Regge-Wheeler, Zerilli and gauge field
  equations}\BibitemShut {NoStop}%
\bibitem [{\citenamefont {Poisson}\ and\ \citenamefont
  {Will}(2014)}]{poisson_will_2014}%
  \BibitemOpen
  \bibfield  {author} {\bibinfo {author} {\bibfnamefont {E.}~\bibnamefont
  {Poisson}}\ and\ \bibinfo {author} {\bibfnamefont {C.~M.}\ \bibnamefont
  {Will}},\ }\bibinfo {title} {Radiative losses and radiation reaction},\ in\
  \href {https://doi.org/10.1017/CBO9781139507486.013} {\emph {\bibinfo
  {booktitle} {Gravity: Newtonian, Post-Newtonian, Relativistic}}}\ (\bibinfo
  {publisher} {Cambridge University Press},\ \bibinfo {year} {2014})\ p.\
  \bibinfo {pages} {624–698}\BibitemShut {NoStop}%
\bibitem [{\citenamefont {Mano}\ \emph {et~al.}(1996)\citenamefont {Mano},
  \citenamefont {Suzuki},\ and\ \citenamefont {Takasugi}}]{Mano:1996vt}%
  \BibitemOpen
  \bibfield  {author} {\bibinfo {author} {\bibfnamefont {S.}~\bibnamefont
  {Mano}}, \bibinfo {author} {\bibfnamefont {H.}~\bibnamefont {Suzuki}},\ and\
  \bibinfo {author} {\bibfnamefont {E.}~\bibnamefont {Takasugi}},\ }\href
  {https://doi.org/10.1143/PTP.95.1079} {\bibfield  {journal} {\bibinfo
  {journal} {Prog. Theor. Phys.}\ }\textbf {\bibinfo {volume} {95}},\ \bibinfo
  {pages} {1079} (\bibinfo {year} {1996})},\ \Eprint
  {https://arxiv.org/abs/gr-qc/9603020} {arXiv:gr-qc/9603020} \BibitemShut
  {NoStop}%
\bibitem [{\citenamefont {Casals}\ and\ \citenamefont
  {Ottewill}(2015)}]{Casals:2015nja}%
  \BibitemOpen
  \bibfield  {author} {\bibinfo {author} {\bibfnamefont {M.}~\bibnamefont
  {Casals}}\ and\ \bibinfo {author} {\bibfnamefont {A.~C.}\ \bibnamefont
  {Ottewill}},\ }\href {https://doi.org/10.1103/PhysRevD.92.124055} {\bibfield
  {journal} {\bibinfo  {journal} {Phys. Rev. D}\ }\textbf {\bibinfo {volume}
  {92}},\ \bibinfo {pages} {124055} (\bibinfo {year} {2015})},\ \Eprint
  {https://arxiv.org/abs/1509.04702} {arXiv:1509.04702 [gr-qc]} \BibitemShut
  {NoStop}%
\bibitem [{\citenamefont {Hopper}\ and\ \citenamefont
  {Evans}(2010)}]{PhysRevD.82.084010}%
  \BibitemOpen
  \bibfield  {author} {\bibinfo {author} {\bibfnamefont {S.}~\bibnamefont
  {Hopper}}\ and\ \bibinfo {author} {\bibfnamefont {C.~R.}\ \bibnamefont
  {Evans}},\ }\href {https://doi.org/10.1103/PhysRevD.82.084010} {\bibfield
  {journal} {\bibinfo  {journal} {Phys. Rev. D}\ }\textbf {\bibinfo {volume}
  {82}},\ \bibinfo {pages} {084010} (\bibinfo {year} {2010})}\BibitemShut
  {NoStop}%
\bibitem [{\citenamefont {Dolan}\ and\ \citenamefont
  {Barack}(2013)}]{Dolan:2012jg}%
  \BibitemOpen
  \bibfield  {author} {\bibinfo {author} {\bibfnamefont {S.~R.}\ \bibnamefont
  {Dolan}}\ and\ \bibinfo {author} {\bibfnamefont {L.}~\bibnamefont {Barack}},\
  }\href {https://doi.org/10.1103/PhysRevD.87.084066} {\bibfield  {journal}
  {\bibinfo  {journal} {Phys. Rev. D}\ }\textbf {\bibinfo {volume} {87}},\
  \bibinfo {pages} {084066} (\bibinfo {year} {2013})},\ \Eprint
  {https://arxiv.org/abs/1211.4586} {arXiv:1211.4586 [gr-qc]} \BibitemShut
  {NoStop}%
\bibitem [{\citenamefont {Isoyama}\ \emph {et~al.}(2014)\citenamefont
  {Isoyama}, \citenamefont {Barack}, \citenamefont {Dolan}, \citenamefont
  {Le~Tiec}, \citenamefont {Nakano}, \citenamefont {Shah}, \citenamefont
  {Tanaka},\ and\ \citenamefont {Warburton}}]{Isoyama:2014mja}%
  \BibitemOpen
  \bibfield  {author} {\bibinfo {author} {\bibfnamefont {S.}~\bibnamefont
  {Isoyama}}, \bibinfo {author} {\bibfnamefont {L.}~\bibnamefont {Barack}},
  \bibinfo {author} {\bibfnamefont {S.~R.}\ \bibnamefont {Dolan}}, \bibinfo
  {author} {\bibfnamefont {A.}~\bibnamefont {Le~Tiec}}, \bibinfo {author}
  {\bibfnamefont {H.}~\bibnamefont {Nakano}}, \bibinfo {author} {\bibfnamefont
  {A.~G.}\ \bibnamefont {Shah}}, \bibinfo {author} {\bibfnamefont
  {T.}~\bibnamefont {Tanaka}},\ and\ \bibinfo {author} {\bibfnamefont
  {N.}~\bibnamefont {Warburton}},\ }\href
  {https://doi.org/10.1103/PhysRevLett.113.161101} {\bibfield  {journal}
  {\bibinfo  {journal} {Phys. Rev. Lett.}\ }\textbf {\bibinfo {volume} {113}},\
  \bibinfo {pages} {161101} (\bibinfo {year} {2014})},\ \Eprint
  {https://arxiv.org/abs/1404.6133} {arXiv:1404.6133 [gr-qc]} \BibitemShut
  {NoStop}%
\bibitem [{\citenamefont {Detweiler}\ and\ \citenamefont
  {Poisson}(2004)}]{PhysRevD.69.084019}%
  \BibitemOpen
  \bibfield  {author} {\bibinfo {author} {\bibfnamefont {S.}~\bibnamefont
  {Detweiler}}\ and\ \bibinfo {author} {\bibfnamefont {E.}~\bibnamefont
  {Poisson}},\ }\href {https://doi.org/10.1103/PhysRevD.69.084019} {\bibfield
  {journal} {\bibinfo  {journal} {Phys. Rev. D}\ }\textbf {\bibinfo {volume}
  {69}},\ \bibinfo {pages} {084019} (\bibinfo {year} {2004})}\BibitemShut
  {NoStop}%
\bibitem [{\citenamefont {Teukolsky}(1973)}]{Teukolsky:1973ha}%
  \BibitemOpen
  \bibfield  {author} {\bibinfo {author} {\bibfnamefont {S.~A.}\ \bibnamefont
  {Teukolsky}},\ }\href {https://doi.org/10.1086/152444} {\bibfield  {journal}
  {\bibinfo  {journal} {Astrophys. J.}\ }\textbf {\bibinfo {volume} {185}},\
  \bibinfo {pages} {635} (\bibinfo {year} {1973})}\BibitemShut {NoStop}%
\bibitem [{\citenamefont {Chrzanowski}(1975)}]{Chrzanowski:1975wv}%
  \BibitemOpen
  \bibfield  {author} {\bibinfo {author} {\bibfnamefont {P.~L.}\ \bibnamefont
  {Chrzanowski}},\ }\href {https://doi.org/10.1103/PhysRevD.11.2042} {\bibfield
   {journal} {\bibinfo  {journal} {Phys. Rev. D}\ }\textbf {\bibinfo {volume}
  {11}},\ \bibinfo {pages} {2042} (\bibinfo {year} {1975})}\BibitemShut
  {NoStop}%
\bibitem [{\citenamefont {Kegeles}\ and\ \citenamefont
  {Cohen}(1979)}]{Kegeles:1979an}%
  \BibitemOpen
  \bibfield  {author} {\bibinfo {author} {\bibfnamefont {L.~S.}\ \bibnamefont
  {Kegeles}}\ and\ \bibinfo {author} {\bibfnamefont {J.~M.}\ \bibnamefont
  {Cohen}},\ }\href {https://doi.org/10.1103/PhysRevD.19.1641} {\bibfield
  {journal} {\bibinfo  {journal} {Phys. Rev. D}\ }\textbf {\bibinfo {volume}
  {19}},\ \bibinfo {pages} {1641} (\bibinfo {year} {1979})}\BibitemShut
  {NoStop}%
\bibitem [{\citenamefont {Wald}(1978)}]{Wald:1978vm}%
  \BibitemOpen
  \bibfield  {author} {\bibinfo {author} {\bibfnamefont {R.~M.}\ \bibnamefont
  {Wald}},\ }\href {https://doi.org/10.1103/PhysRevLett.41.203} {\bibfield
  {journal} {\bibinfo  {journal} {Phys. Rev. Lett.}\ }\textbf {\bibinfo
  {volume} {41}},\ \bibinfo {pages} {203} (\bibinfo {year} {1978})}\BibitemShut
  {NoStop}%
\bibitem [{\citenamefont {Green}\ \emph {et~al.}(2020)\citenamefont {Green},
  \citenamefont {Hollands},\ and\ \citenamefont {Zimmerman}}]{Green:2019nam}%
  \BibitemOpen
  \bibfield  {author} {\bibinfo {author} {\bibfnamefont {S.~R.}\ \bibnamefont
  {Green}}, \bibinfo {author} {\bibfnamefont {S.}~\bibnamefont {Hollands}},\
  and\ \bibinfo {author} {\bibfnamefont {P.}~\bibnamefont {Zimmerman}},\ }\href
  {https://doi.org/10.1088/1361-6382/ab7075} {\bibfield  {journal} {\bibinfo
  {journal} {Class. Quant. Grav.}\ }\textbf {\bibinfo {volume} {37}},\ \bibinfo
  {pages} {075001} (\bibinfo {year} {2020})},\ \Eprint
  {https://arxiv.org/abs/1908.09095} {arXiv:1908.09095 [gr-qc]} \BibitemShut
  {NoStop}%
\bibitem [{\citenamefont {Toomani}\ \emph {et~al.}(2022)\citenamefont
  {Toomani}, \citenamefont {Zimmerman}, \citenamefont {Spiers}, \citenamefont
  {Hollands}, \citenamefont {Pound},\ and\ \citenamefont
  {Green}}]{Toomani:2021jlo}%
  \BibitemOpen
  \bibfield  {author} {\bibinfo {author} {\bibfnamefont {V.}~\bibnamefont
  {Toomani}}, \bibinfo {author} {\bibfnamefont {P.}~\bibnamefont {Zimmerman}},
  \bibinfo {author} {\bibfnamefont {A.}~\bibnamefont {Spiers}}, \bibinfo
  {author} {\bibfnamefont {S.}~\bibnamefont {Hollands}}, \bibinfo {author}
  {\bibfnamefont {A.}~\bibnamefont {Pound}},\ and\ \bibinfo {author}
  {\bibfnamefont {S.~R.}\ \bibnamefont {Green}},\ }\href
  {https://doi.org/10.1088/1361-6382/ac37a5} {\bibfield  {journal} {\bibinfo
  {journal} {Class. Quant. Grav.}\ }\textbf {\bibinfo {volume} {39}},\ \bibinfo
  {pages} {015019} (\bibinfo {year} {2022})},\ \Eprint
  {https://arxiv.org/abs/2108.04273} {arXiv:2108.04273 [gr-qc]} \BibitemShut
  {NoStop}%
\bibitem [{\citenamefont {Sundararajan}\ \emph {et~al.}(2007)\citenamefont
  {Sundararajan}, \citenamefont {Khanna},\ and\ \citenamefont
  {Hughes}}]{Sundararajan:2007jg}%
  \BibitemOpen
  \bibfield  {author} {\bibinfo {author} {\bibfnamefont {P.~A.}\ \bibnamefont
  {Sundararajan}}, \bibinfo {author} {\bibfnamefont {G.}~\bibnamefont
  {Khanna}},\ and\ \bibinfo {author} {\bibfnamefont {S.~A.}\ \bibnamefont
  {Hughes}},\ }\href {https://doi.org/10.1103/PhysRevD.76.104005} {\bibfield
  {journal} {\bibinfo  {journal} {Phys. Rev. D}\ }\textbf {\bibinfo {volume}
  {76}},\ \bibinfo {pages} {104005} (\bibinfo {year} {2007})},\ \Eprint
  {https://arxiv.org/abs/gr-qc/0703028} {arXiv:gr-qc/0703028} \BibitemShut
  {NoStop}%
\bibitem [{\citenamefont {Harms}\ \emph {et~al.}(2014)\citenamefont {Harms},
  \citenamefont {Bernuzzi}, \citenamefont {Nagar},\ and\ \citenamefont
  {Zenginoglu}}]{Harms:2014dqa}%
  \BibitemOpen
  \bibfield  {author} {\bibinfo {author} {\bibfnamefont {E.}~\bibnamefont
  {Harms}}, \bibinfo {author} {\bibfnamefont {S.}~\bibnamefont {Bernuzzi}},
  \bibinfo {author} {\bibfnamefont {A.}~\bibnamefont {Nagar}},\ and\ \bibinfo
  {author} {\bibfnamefont {A.}~\bibnamefont {Zenginoglu}},\ }\href
  {https://doi.org/10.1088/0264-9381/31/24/245004} {\bibfield  {journal}
  {\bibinfo  {journal} {Class. Quant. Grav.}\ }\textbf {\bibinfo {volume}
  {31}},\ \bibinfo {pages} {245004} (\bibinfo {year} {2014})},\ \Eprint
  {https://arxiv.org/abs/1406.5983} {arXiv:1406.5983 [gr-qc]} \BibitemShut
  {NoStop}%
\bibitem [{\citenamefont {Heffernan}\ \emph {et~al.}(2018)\citenamefont
  {Heffernan}, \citenamefont {Ottewill}, \citenamefont {Warburton},
  \citenamefont {Wardell},\ and\ \citenamefont {Diener}}]{Heffernan:2017cad}%
  \BibitemOpen
  \bibfield  {author} {\bibinfo {author} {\bibfnamefont {A.}~\bibnamefont
  {Heffernan}}, \bibinfo {author} {\bibfnamefont {A.~C.}\ \bibnamefont
  {Ottewill}}, \bibinfo {author} {\bibfnamefont {N.}~\bibnamefont {Warburton}},
  \bibinfo {author} {\bibfnamefont {B.}~\bibnamefont {Wardell}},\ and\ \bibinfo
  {author} {\bibfnamefont {P.}~\bibnamefont {Diener}},\ }\href
  {https://doi.org/10.1088/1361-6382/aad420} {\bibfield  {journal} {\bibinfo
  {journal} {Class. Quant. Grav.}\ }\textbf {\bibinfo {volume} {35}},\ \bibinfo
  {pages} {194001} (\bibinfo {year} {2018})},\ \Eprint
  {https://arxiv.org/abs/1712.01098} {arXiv:1712.01098 [gr-qc]} \BibitemShut
  {NoStop}%
\bibitem [{\citenamefont {Fujita}(2012)}]{Fujita:2012cm}%
  \BibitemOpen
  \bibfield  {author} {\bibinfo {author} {\bibfnamefont {R.}~\bibnamefont
  {Fujita}},\ }\href {https://doi.org/10.1143/PTP.128.971} {\bibfield
  {journal} {\bibinfo  {journal} {Prog. Theor. Phys.}\ }\textbf {\bibinfo
  {volume} {128}},\ \bibinfo {pages} {971} (\bibinfo {year} {2012})},\ \Eprint
  {https://arxiv.org/abs/1211.5535} {arXiv:1211.5535 [gr-qc]} \BibitemShut
  {NoStop}%
\bibitem [{\citenamefont {Kavanagh}\ \emph {et~al.}(2015)\citenamefont
  {Kavanagh}, \citenamefont {Ottewill},\ and\ \citenamefont
  {Wardell}}]{Kavanagh:2015lva}%
  \BibitemOpen
  \bibfield  {author} {\bibinfo {author} {\bibfnamefont {C.}~\bibnamefont
  {Kavanagh}}, \bibinfo {author} {\bibfnamefont {A.~C.}\ \bibnamefont
  {Ottewill}},\ and\ \bibinfo {author} {\bibfnamefont {B.}~\bibnamefont
  {Wardell}},\ }\href {https://doi.org/10.1103/PhysRevD.92.084025} {\bibfield
  {journal} {\bibinfo  {journal} {Phys. Rev. D}\ }\textbf {\bibinfo {volume}
  {92}},\ \bibinfo {pages} {084025} (\bibinfo {year} {2015})},\ \Eprint
  {https://arxiv.org/abs/1503.02334} {arXiv:1503.02334 [gr-qc]} \BibitemShut
  {NoStop}%
\bibitem [{\citenamefont {Bini}\ and\ \citenamefont
  {Damour}(2014)}]{Bini:2014ica}%
  \BibitemOpen
  \bibfield  {author} {\bibinfo {author} {\bibfnamefont {D.}~\bibnamefont
  {Bini}}\ and\ \bibinfo {author} {\bibfnamefont {T.}~\bibnamefont {Damour}},\
  }\href {https://doi.org/10.1103/PhysRevD.90.024039} {\bibfield  {journal}
  {\bibinfo  {journal} {Phys. Rev. D}\ }\textbf {\bibinfo {volume} {90}},\
  \bibinfo {pages} {024039} (\bibinfo {year} {2014})},\ \Eprint
  {https://arxiv.org/abs/1404.2747} {arXiv:1404.2747 [gr-qc]} \BibitemShut
  {NoStop}%
\bibitem [{\citenamefont {Munna}\ and\ \citenamefont
  {Evans}(2022)}]{Munna:2022xts}%
  \BibitemOpen
  \bibfield  {author} {\bibinfo {author} {\bibfnamefont {C.}~\bibnamefont
  {Munna}}\ and\ \bibinfo {author} {\bibfnamefont {C.~R.}\ \bibnamefont
  {Evans}},\ }\href {https://doi.org/10.1103/PhysRevD.106.044058} {\bibfield
  {journal} {\bibinfo  {journal} {Phys. Rev. D}\ }\textbf {\bibinfo {volume}
  {106}},\ \bibinfo {pages} {044058} (\bibinfo {year} {2022})},\ \Eprint
  {https://arxiv.org/abs/2206.04085} {arXiv:2206.04085 [gr-qc]} \BibitemShut
  {NoStop}%
\bibitem [{\citenamefont {Kavanagh}(2021)}]{Kavanagh:2021}%
  \BibitemOpen
  \bibfield  {author} {\bibinfo {author} {\bibfnamefont {C.}~\bibnamefont
  {Kavanagh}}} (\bibinfo {year} {2021}),\ \bibinfo {note} {the 24th Capra
  meeting on radiation reaction in general relativity}\BibitemShut {NoStop}%
\end{thebibliography}%
	
\end{document}